\pdfoutput=1
\documentclass[5p,twocolumn,a4paper,times,12pt]{elsarticle}

\usepackage{graphicx}
\usepackage{amssymb}
\usepackage{wasysym}   
\usepackage{hyperref}
\usepackage{makecell}

\journal{Astroparticle Physics}

\newcommand{\HADRON}{HAD\-RON }
\newcommand{\GAMMA}{GAM\-MA }
\newcommand{\KASCADE}{KAS\-CA\-DE }
\newcommand{\idest}{\textit{i.\,e. }}
\newcommand{\eg}{\textit{e.\,g. }}

\begin{document}

\begin{frontmatter}
\title{The rise of muon content in extensive air showers after the~3\,PeV~knee of the~primary cosmic ray spectrum according to~data of the~Tien~Shan mountain installation}

\author[lpi]{A.\,L.\,Shepetov}
\ead{ashep@tien-shan.org} 

\author[lpi]{S.\,B.\,Shaulov}

\author[lpi]{O.\,I.\,Likiy}

\author[lpi]{V.\,A.\,Ryabov}

\author[kaz1]{T.\,Kh.\,Sadykov}
\author[kaz2]{N.\,O.\,Saduev}

\author[lpi]{V.\,V.\,Zhukov}

\address[lpi]{P.\,N.\,Lebedev Physical Institute of the Russian Academy of Sciences (LPI),  119991, Leninsky pr., 53, Moscow, Russia}

\address[kaz1]{Satbayev University, Institute of Physics and Technology, 050032, Ibragimova str. 11, Almaty, Kazakhstan}

\address[kaz2]{Al-Farabi Kazakh National University, Institute of Experimental and Theoretical Physics, 050040, Al-Farabi av., 71, Almaty, Kazakhstan}


\begin{abstract}
We put together the experimental results on muon component of extensive air showers (EAS) which were gained with various techniques at the detector complex of the Tien Shan mountain station. According to this comparison, the problem of the EAS muon content in the range of primary cosmic ray energies $(1-100)$\,PeV seems to be more complicated than it was usually supposed. Generally, from the models of nuclear interaction it follows that the EAS which have produced gamma-hadron families in the Tien Shan X-ray emulsion chamber did preferably originate from interaction of the light cosmic ray nuclei, such that their muon abundance must be $\sim$$1.5$~times below an average calculated over all showers. In contrary, the experimental muon counts in the EAS with families demonstrate a $(1.5-2)$-fold excess above the average, and this difference starts to be observable in the showers with the energy above the $3$\,PeV knee of the primary cosmic ray spectrum. Later on, the rise of muon production in EAS after the knee was confirmed at Tien Shan by another experiment on detection of the neutrons stemmed from interaction of cosmic ray muons. Thus, the results obtained by the two completely different methods do mutually agree with each other but contradict to the common models of hadron interaction.
\end{abstract}


\end{frontmatter}

\section{Introduction}

Due to the unique penetrative capability of muons many still obscure problems of the physics of cosmic rays can be elucidated by the study of their muonic component. It is known that at first interaction of a high energy cosmic ray particle  which generally takes place somewhere at a height of (20--30)\,km in the Earth's atmosphere, a multitude of pions and kaons is born whose further fate determines development of a subsequent extensive air shower (EAS). Secondary mesons can either collide in their turn with atmosphere nuclei giving rise to the next generation of nuclear-active particles, or to decay by one of the weak interaction channels. In latter case a multitude of muons is originated which can reach a ground-based detector without any subsequent interaction. Such muons is a sole informational source for direct study of high-energy nuclear reactions which proceed in the very beginning of the EAS cascade process, practically at the energy of primary cosmic ray particle.

Since the muons detected at the ground level arise as interaction result of cosmic ray hadrons, the phenomenological characteristics of muon signal in EAS, and primarily the mean number of muons $N_\mu$ in a shower, depend strongly on the fine details of elementary interaction model, such as the inelasticity coefficient, the charge ratio, the multiplicity and  production probability of the baryon-antibaryon pairs, \textit{etc} \cite{muons3,muons1,crvslhc_dembinski_2019}. On the other hand, the muon content of EAS is obviously sensitive to the mass composition of the cosmic ray nuclei \cite{muonseasmass_2004,2009auger,muons2,muonseas_2015}, since the average amount of muons produced at one and the same interaction energy grows in accordance with the atomic mass $A$ of interacting particles as $N_\mu\sim A^\alpha$, where $\alpha\approx(0.1-0.2)$. 
Advanced study of the properties of muon signal, especially in combination with other EAS observables, may be an instrument to disentangle the interconnection between the different factors which affect the muon abundance in EAS.

Investigation of the muon component of EAS has remained one of the main directions of experimental activity at the Tien Shan mountain cosmic ray station of LPI over the whole history of its existence. During the four last decades such studies were made here in the frames of two absolutely different experiments: in the years 1985--1991, at the muon hodoscope which was a part of the complex installation \HADRON \cite{shaulov_icrc1991,ontienhistory1,epj2017_shaulov}, and later on, since the beginning of 2000s and up to the present time, when registration of the signals of muon passage was continued there at an underground neutron detector \cite{undgour2008icrcmexico,undgour1,undgour2,our2018_undg}. Since the year 2015 the underground detector operates together with a system of shower particles detectors \cite{ontien-nim2016} specially for the investigation purpose of the muon component of EAS.

\begin{table*}
\begin{center}
\caption{Composition of the Tien Shan experimental complex in the various periods of its research activity.}
\label{tabepochs}
\begin{tabular*}{0.8\textwidth}{@{\extracolsep{\fill}}c|c|c}
\hline
\hline
detector subsystem&
\makecell{1985--1991\\(\HADRON experiment)}&
2015--present time\\
\hline
\hline
\makecell{the system\\of shower particles\\detectors}&
{\Large +}&
{\Large +}\\

\hline
\makecell{XREC\\and ionization\\chambers}&
{\Large +}&
{\Large--}\\

\hline
\makecell{the muon\\hodoscope}&
{\Large +}&
{\Large--}\\

\hline
\makecell{the underground\\neutron detector\\for registration of\\muon interactions}&
{\Large--}&
{\Large +}
\\

\hline
\hline
\end{tabular*}
\end{center}
\end{table*}

The configurations of the detector complex at the Tien Shan mountain station in the various periods of time are listed in Table~\ref{tabepochs}. All measurements were made at a height of $3340$\,m above the sea level.

The subject of present publication is to compare the old results of the \HADRON installation on the muonic accompaniment of EAS against the data which were newly gained with the underground neutron detector, and to demonstrate a remarkable qualitative parallelism between such diverse experiments in what concerns peculiar behavior of EAS muons in the range of primary cosmic ray energy of $(1-100)$\,PeV.

Correspondingly, the structure of the article is as following.  After a review  made in Section~\ref{secinst} of the experimental facilities used at Tien~Shan for detection of the different cosmic ray components essential for the present study, in the next two sections the results are discussed concerning the phenomenological properties of EAS muons which arise from the two datasets: the data which were obtained at the time of \HADRON experiment are considered in Section~\ref{sechadron}, while Section~\ref{secundg} relates to the studies at the underground neutron detector in present time. The final remarks on the properties of the EAS muon component which follow from the results of both independent experiments are summed up in Conclusion section.

%
%
%
%
%


\section{Instrumentation}
\label{secinst}

\subsection{The complex hadron detector of the Tien Shan mountain station: X-ray films and ionization chambers}

\begin{figure*}
{\centering
\includegraphics[width=0.52\textwidth, trim=0mm 60mm 0mm 111mm]{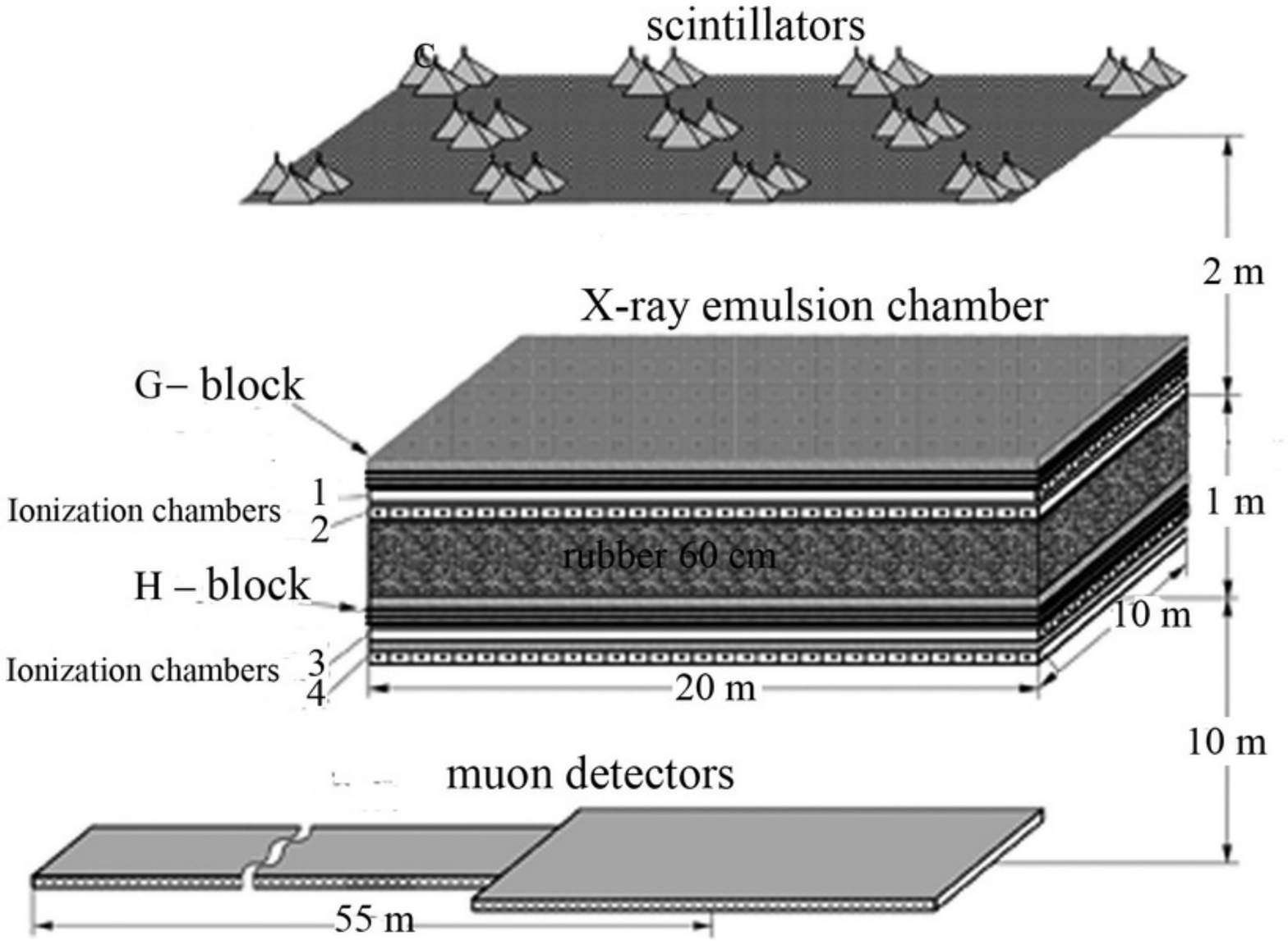}
\includegraphics[width=0.47\textwidth, trim=10mm 33mm 120mm 0mm]{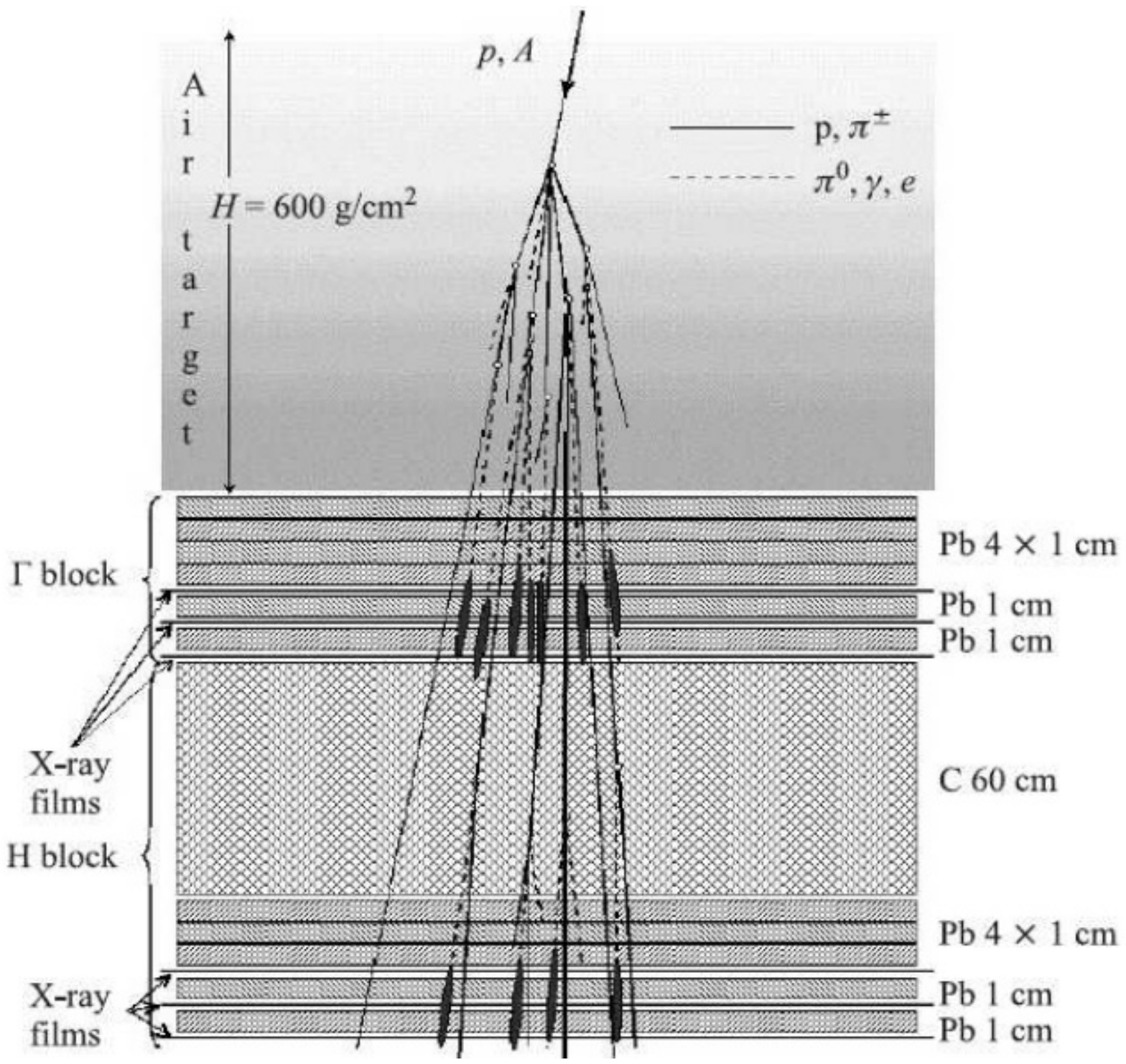}
\caption{The scheme of detector disposition in the central part of the Tien~Shan detector complex at the time of \HADRON experiment (1985--1991). Left frame, from top to bottom: the ''carpet'' of the charged  particles detectors; the hadron detector (X-ray emulsion films and ionization chambers); the muon hodoscope. Right frame: the cut-off and operation principle illustration of a two-storied X-ray emulsion chamber.
}
\label{figihadro}}
\end{figure*}

The \HADRON experiment which was running at the Tien Shan station in the years of 1985--1991 used a hybrid set-up where the electronic instruments of cosmic ray investigation, such as a system of charged particles detectors for registration of EAS passages, and ionization chambers for detection of the integral energy release of hadron component of showers, were combined with the X-ray emulsion technique. General scheme of the central part of \HADRON installation specially aimed for the study of hadronic interactions in the core region of EAS is shown in  Fig.\,\ref{figihadro}.

The operation principle of the two types of hadron detectors applied at the Tien Shan station, both the ionization and X-ray emulsion chambers, was based on conversion of the energy of cosmic ray hadrons into electromagnetic signal of electron-photon cascades developing in a dense medium which was the heavy absorber of the detector. In the \HADRON installation at Tien Shan there were two such absorber parts: the upper gamma-block ($G$-block in Fig.\,\ref{figihadro}), and the lower $H$-block. Both absorbers were composed of the
lead plates alternating with X-ray films, such that their total thickness was of about one interaction length for the high energy hadrons, or of $(8-10)$ electromagnetic cascade units. An additional target for interaction of cosmic ray hadrons which consisted of a 60\,cm thick layer of rubber enriched with carbon ((CH)$_n$) separated the blocks in vertical direction \cite{ontientu}.

The continuous sensitive area of the central hadron detector in the \HADRON installation was 160\,m$^2$.

During the \HADRON experiment the high energy gamma rays produced in decay of neutral mesons which had originated from interaction of cosmic ray hadrons in the atmosphere above the installation were giving rise to electron-photon cascades developing in the lead absorber of its upper ($G$) block. In the lower $H$-block such cascades were initiated by the gamma rays from decay of pions which were born within the carbon target by those cosmic ray particles which had slipped through the atmosphere and the upper block without interaction.


Ionization created by electron-photon cascades in the hadron detector was collected by the gas discharge ionization chambers made of a $(0.25\times 0.12 \times 200)$\,cm$^3$ copper profile and filled with technical argon. There were two sets of such chambers both in the $G$- and $H$-block, with their long axes laid crosswise to ensure rough estimation of the place of cascade development in a $2$-dimensional coordinate frame.

Besides an electrical signal on the anode wires of ionization chambers, the cascades initiated by the gamma ray quanta with sufficient energy, of about 1\,TeV and higher, can cause darkening of X-ray emulsion in the films which were put between the sheets of lead absorber in the $G$- and $H$-blocks. After development of films, the optical density of a darkened spot, and consequently the  density of the electron flux in cascade and the energy of a primary gamma ray quantum can be estimated through photo-densitometry of the spots. This effect was laid in the basis of the X-ray emulsion chamber (XREC) investigation method of the properties of high energy hadron interaction in cosmic rays. Thus, the real time electronic method of cascade detection by the means of ionization chambers was applied for target designation in the XREC, while the latter ensured, after development of X-ray films, precise study of the most energetic
EAS hadrons with unprecedented spatial, $\lesssim$$100$\,$\mu$m, and energy, $\sim$$(20-30)$\%, resolution. 

The energy threshold of hadron detection at the Tien~Shan station was of about 100\,GeV in the channels of ionization chamber, and of $\sim$(1--2)\,TeV for the XREC. Generally, the hadrons with such energies are concentrated in the central region of EAS, so the Tien Shan complex detector was a mean for precise investigation of the structure of high-energy hadron core of extensive air showers.

An excellent spatial resolution of the technique of X-ray films permitted to detect separately the electron-photon cascades caused by the tight groups of genetically connected gamma rays (in $G$-block) and hadrons (in $H$-block) which had emerged from interaction of one and the same primary particle, either a cosmic ray proton or nucleus which had the energy $E_0$ of a few PeV order or above. In the scientific literature on XREC such groups of electromagnetic cascades are commonly referenced to as ``families''. It is the investigation of the properties of gamma ray and hadron families in their dependence on the characteristics of accompanying EAS, and first of all on the energy $E_0$ of primary cosmic ray particle, which was the main goal of the \HADRON experiment.

\subsection{Detection of EAS particles}
\label{sectishw}

In all Tien Shan experiments the electron component of extensive air showers was studied by the means of the charged particles detectors made of solid blocks of polystyrene scintillator and settled nearly equidistantly over an area of several hundreds of square meters order. In the case of an EAS event the sum amplitude of scintillation flash generated in every detector is proportional to the amount of charged particles which have passed in the point of detector location, so registration of the amplitudes in a system of synchronously operating detectors gives spatial distribution of the particles density in the shower. 

As illustrated by Fig.\,\ref{figihadro}, in the years of 1985--1991, when the \HADRON experiment was active, the particles detectors were grouped by three in $11$~points densely distributed over a $(12\times32)$\,m$^2$ area of ''scintillation carpet'' which was situated just above the X-ray films and ionization chambers. Some another detector groups were installed along circumferences which surrounded the central ``carpet'' at distances of $(20-80)$\,m. The particles density estimates in each point at that time were calculated by indications averaged over the three scintillators in each group \cite{ontienhistory1,ontien_icrc1987___through_ne_ru}. Since the year 2015 and up to the present, $72$~single detectors are placed in the nodes of a $3$\,m\,$\times$\,$4$\,m rectangular network distributed over the $(27\times32)$\,m$^2$ ''carpet'' area 
\cite{ontien-nim2016}.

In contrast to X-ray films, the electronic methods of the scintillation signal registration permit to detect the EAS events in real time, so each shower was supplied with its precise timestamp. During the measurements the moment of shower passage was indicated by a special trigger signal which was elaborated in the cases when the total amplitude of scintillation pulses summed over all detectors of the shower installation exceeded a predefined threshold. This trigger initiated simultaneous sampling of the momentary signal amplitudes in the particles detectors and ionization chambers, as well as registration of the signals of the underground muon detector (see below).

\begin{figure*}
{\centering
\includegraphics[width=0.49\textwidth, clip, trim=12mm 10mm 0mm 0mm]{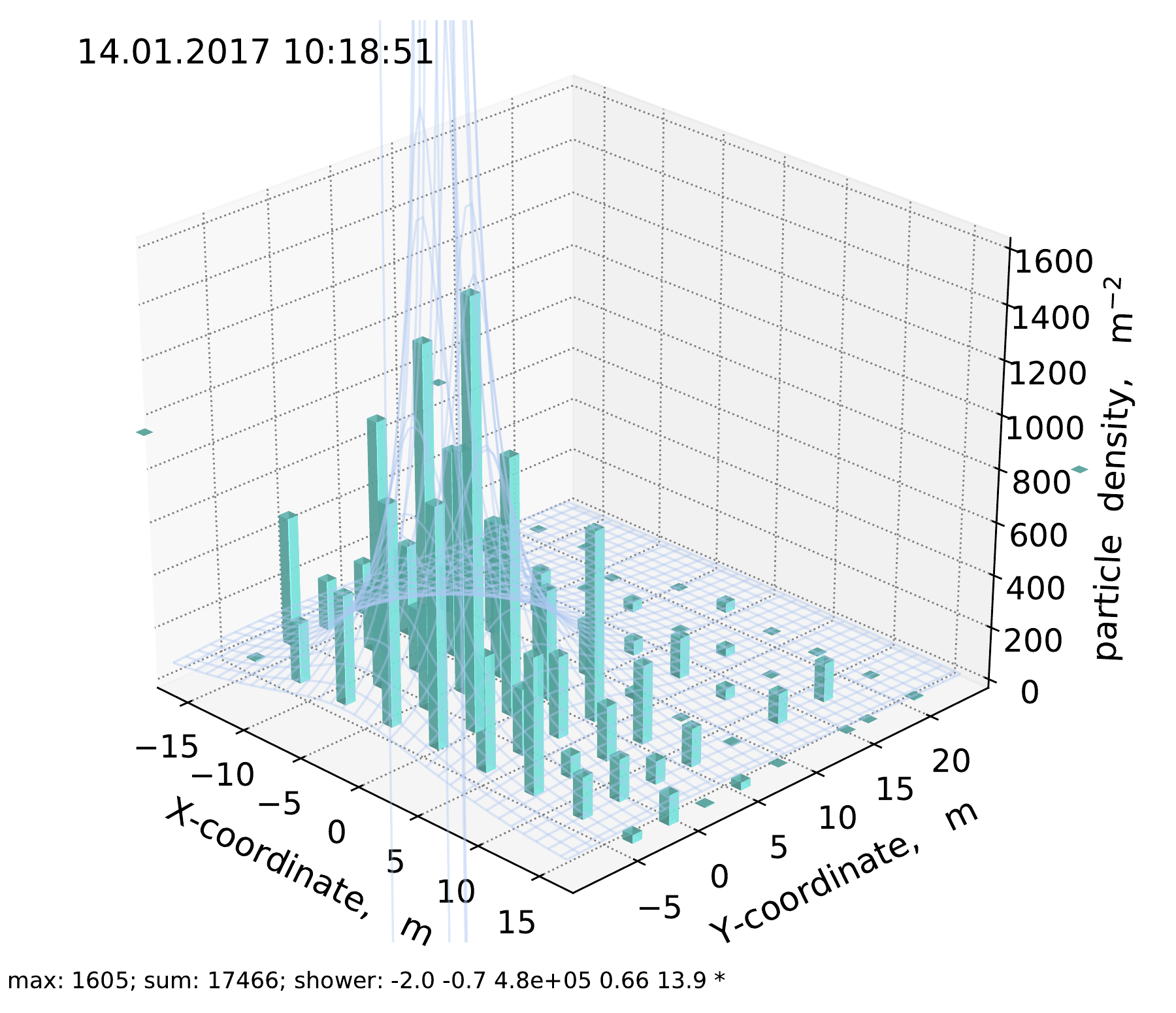}
\includegraphics[width=0.49\textwidth, clip, trim=12mm 10mm 0mm 0mm]{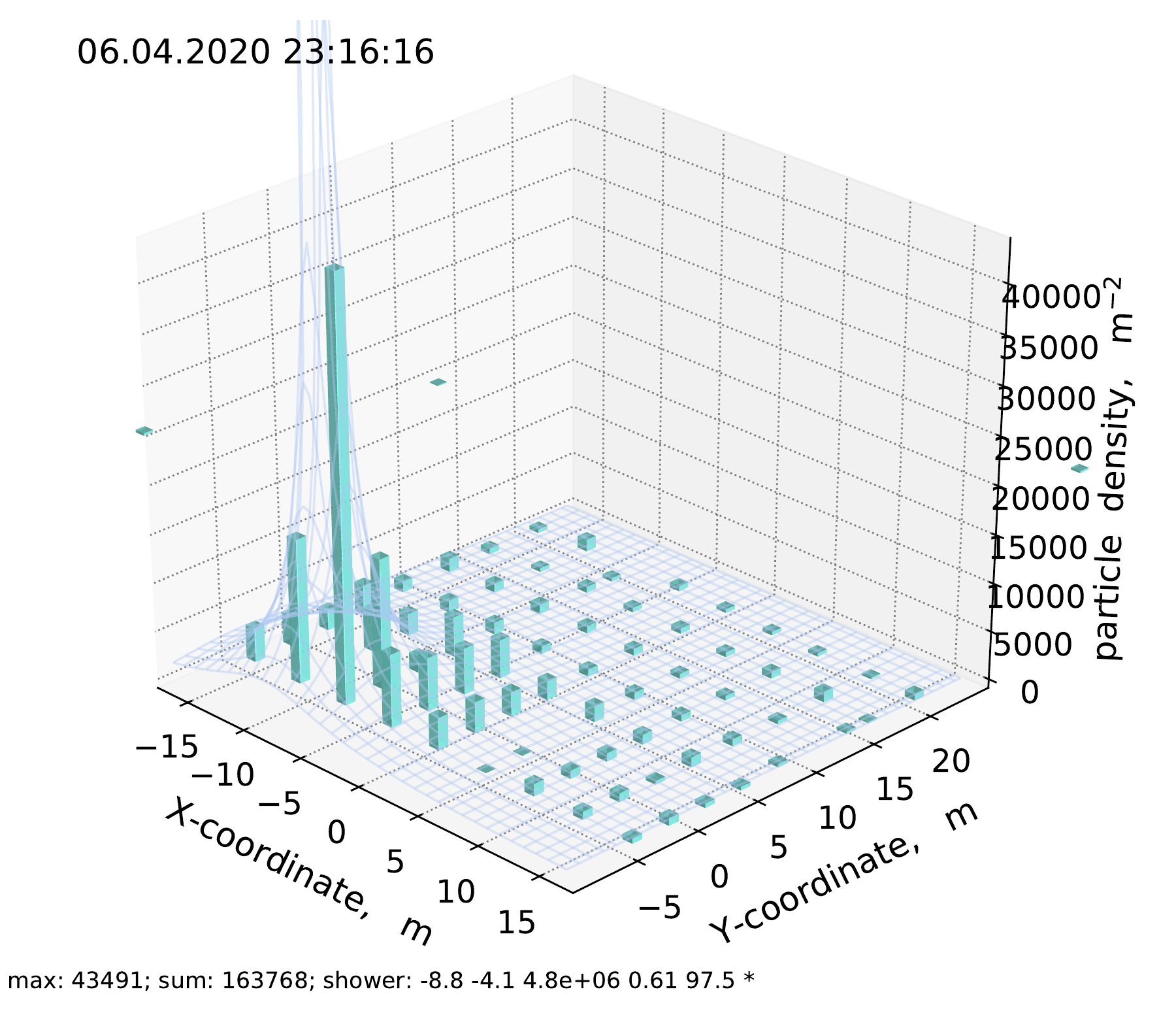}
\caption{Example of extensive air shower events as detected at the Tien Shan installation: two-dimensional spatial distribution of the charged particles density in the central detector ''carpet'' (bars) approximated by a NKG type function (wireframe surface). The primary EAS energy $E_0$ and the shower ``age'' parameter $s$ in the cases presented in the left and right frames were of about $E_0\approx 1.2$\,PeV and $s\approx 0.66$, and $E_0\approx 12$\,PeV and $s\approx 0.61$ correspondingly.}
\label{figieassample}}
\end{figure*}

Mathematical operation of the particles density data registered by the system of scintillation detectors consisted of determination of the basic characteristics of EAS: the pair $(x, y)$ of the coordinates of shower axis, the zenith and azimuth angles of axis direction $(\theta,\varphi)$, the ''age'' parameter $s$, and the estimate for the total number of particles in  shower---the EAS ''size'' $N_e$. This task was solved by fitting the two-dimensional experimentally measured distribution of particles density $\rho_{exp}$ by a function of Nishimura-Kamata-Greisen family (NKG, \cite{nkg-nk}):
\[
\begin{array}{r}
\rho_{NKG}( r, s, N_e )=
0.366 s^2 ( 2.07-s )^{1.25}\times \\
( r / r_M )^{s - 2 }( 1 + r / r_M )^{s - 4.5} / r_M^2 )\times N_e,
\end{array}
\]
$$
\sum\frac{(\rho_{NKG}( r(x,y), s, N_e )-\rho_{exp})^2}{\sigma_\rho^2}\rightarrow \min_{x,y,s,N_e},
$$
where $\sigma_\rho$ is the measurement error of particles density. The Moli\'{e}r radius parameter $r_M$ which appears in the NKG function was accepted to be equal to 120\,m at the altitude of the Tien~Shan station. The detailed algorithms of the particles density calculation over the data of scintillation detectors and succeeding fitting of shower parameters were explained in \cite{ontien-nim2016}. It should be stressed that the applied estimation method of shower parameters is based exclusively on the measurement data of the particles detector system, and does not depend on any specific model of EAS development.

A sample of EAS events as they were seen in the detectors of the Tien Shan shower installation is shown in Fig.\,\ref{figieassample}.

Uncertainties of the shower parameters estimation in the \HADRON experiment were deduced from a series of simulation studies. Of them, the work \cite{2007aragats_simul_like_tien} could be mentioned which was made somewhat later, but using contemporary simulation codes and models of particles interaction, for the EAS installation \GAMMA at mount Aragats. The latter was a total functional analog of the Tien Shan experiment, build on the particles detector devices of the same type, and situated at the same altitude above the sea level. This simulation has shown that the relative accuracy $\Delta N_e/N_e$ of EAS size parameters restored by the density distribution of shower particles among the detectors of such installation is of about $(5-10)$\%, while the mean quadratic error of EAS axis position equals to $(3-5)$\,m.

At the time of \HADRON experiment an alternative possibility existed to pinpoint the location of shower center with a much better precision of $\pm 0.25$\,m, and independently of EAS particles data, by the maximum of amplitude distribution among the ionization chambers of the central hadron detector.

With the known size $N_e$ of an extensive air shower detected at the Tien Shan station, the energy of its primary particle $E_0$ can be evaluated as $E_0(N_e)\approx (2.5\cdot 10^{-6} \times N_e)$\,PeV, with an average relative error of $(25-30)$\%. In contrast to $N_e$ estimates, this relation is model-dependent, and was deduced from a number of EAS development simulations specially made for the Tien Shan installation, \eg \cite{tamada_simul_1994_e0_from_ne}.

In particular, from the latter formula for $E_0(N_e)$ it follows that the shower size value $N_e\approx 10^6$ at the altitude of the Tien Shan station corresponds to EAS created by the particles which belong to the knee region of the primary cosmic ray spectrum at 3\,PeV. As well, the values of EAS energy labelled along the secondary abscissa axes of some experimental data plots which will be presented below were defined in accordance with this re-calculation rule.

\subsection{The muon hodoscope}
\label{secmuohodo}

At the time of \HADRON experiment the registration of the muon component of extensive air showers was made by the means of an underground hodoscope detector built on the basis of Geiger-M\"{u}ller counter tubes.
The hodoscope consisted of $1440$ separate tubes which were divided between $18$ groups, $80$ tubes in each. Every pair of tubes in a group constituted a single informational channel of the hodoscope, with the common area of effective particles detection $\sigma=(2\times 0.062)$\,m$^2$.

\begin{figure*}
{\centering
\includegraphics[width=0.35\textwidth, trim=20mm 0mm 30mm 66mm]{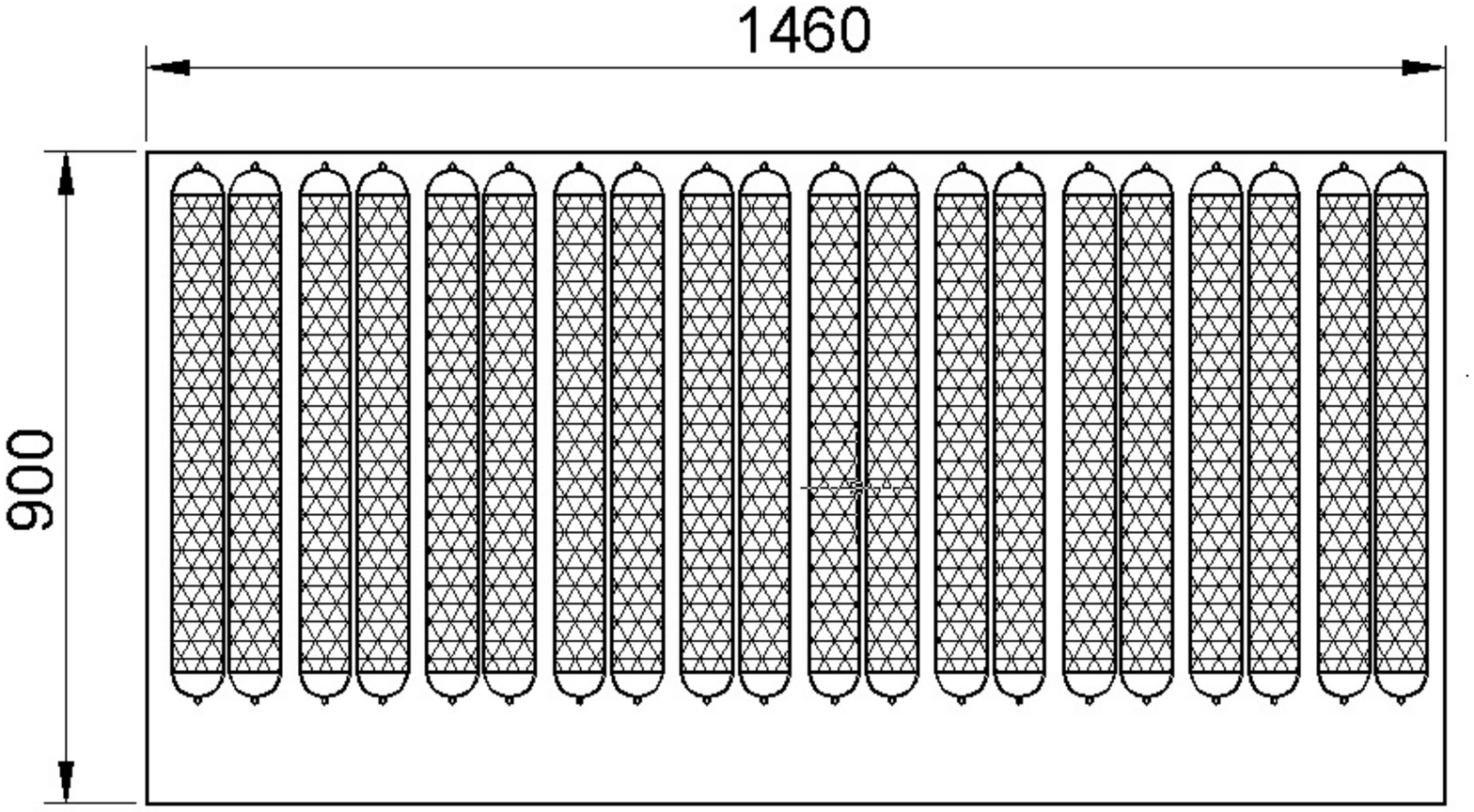}
\includegraphics[width=0.64\textwidth, trim=30mm 44mm 15mm 88mm]{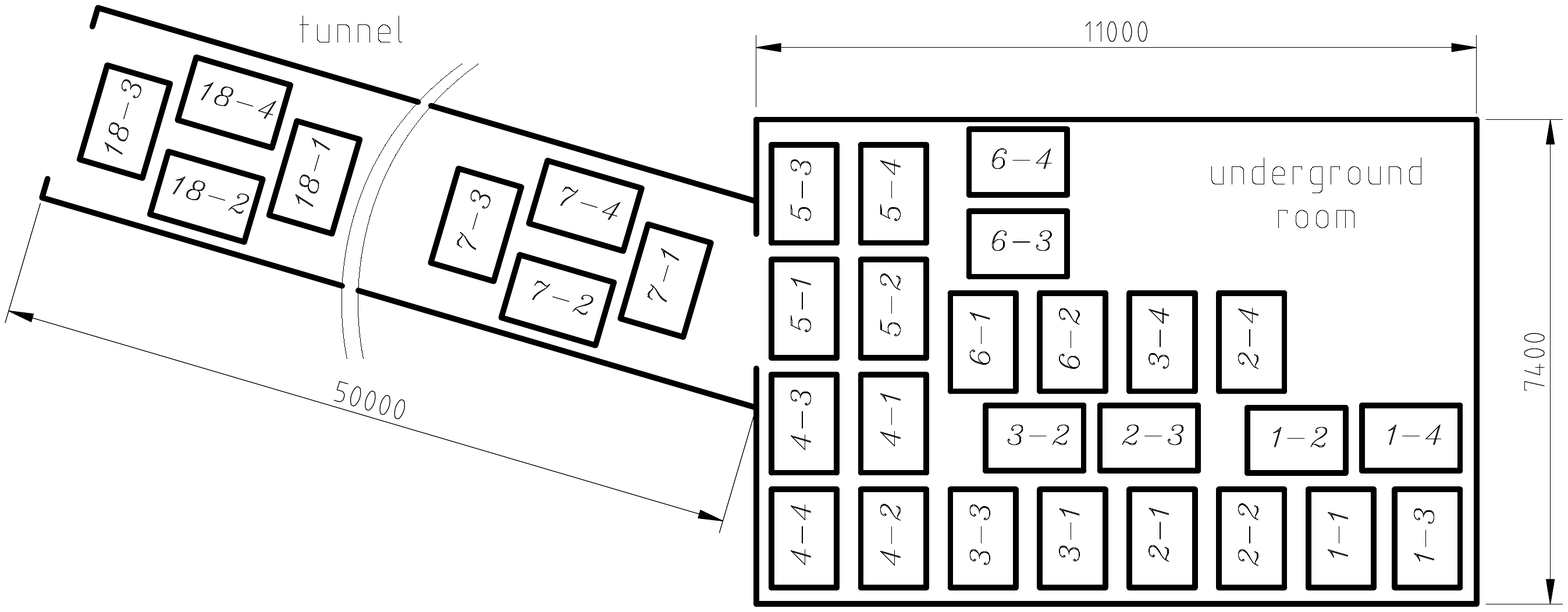}
\caption{The standard detector module of 20~Geiger-M\"{u}ller counters (left), and the disposition scheme of the muon hodoscope modules in the underground room and in the tunnel of the Tien~Shan mountain station at the time of \HADRON experiment. All dimensions are indicated in millimeters.}
\label{figihadroundg}}
\end{figure*}

As it is shown in the placement scheme of Fig.\,\ref{figihadroundg}, six hodoscope groups were settled in an underground room of the Tien Shan station, immediately underneath the X-ray emulsion chamber and the central ''carpet'' of EAS particles detectors, another twelve were distributed over a 50\,m long tunnel which adjoins to that room. Geometrical sizes of the whole space overridden by the central part of the hodoscope just under the center of shower installation and XREC were $(11\times 7)$\,m$^2$, and the sum sensitive area of all counter tubes installed there was of about $30$\,m$^2$.

The thickness of the ground absorber above the underground room and tunnel is of about $2000$\,g/cm$^2$ which corresponds to the energy cut-off of $5$\,GeV for the muons moving in nearly vertical direction. Because of the rapidly falling energy spectrum of cosmic ray particles, the majority of muons detected by the Tien Shan hodoscope had the energies close to this lower limit of a few GeV order.

At the moment of an EAS signaled by the trigger from the system of shower particles detectors the control electronics kept for each hodoscope group the numbers $m$ of its informational channels in which the tubes were fired by passage of EAS muons. Afterwards the most probable value for the density of muon flux in the disposition point of the group can be defined as
$$
\rho=\frac{1}{\sigma}\ln\left(\frac{n}{n-m}\right),
$$
where $n=40$ is the total number of channels in a group, and $\sigma=0.12$\,m$^2$ is the sum sensitive area of two counter tubes in a channel. Since the central region of each EAS just beneath the X-ray emulsion chamber was of most interest in the \HADRON experiment, the average muon density in the underground room at that times was calculated as
$$
\overline\rho_{undg}=\frac{1}{\sigma}\ln
\left(\frac{k\cdot40}{k\cdot40-\sum_{i=1}^{k}m_{i}}\right),
$$
where $k\leqslant 6$ is the amount of hodoscope groups which were operating in the underground room at the moment of EAS detection, and $m_i$ is the number of the channels fired in each group. The local density values of the muon flux at EAS periphery were defined analogically, using the data of hodoscope groups hosted along the underground tunnel.

With known $\overline\rho_{undg}$, the total number of muons in shower $N_\mu$ was estimated using the average lateral distribution function of the muon flux in EAS, $\varphi_{\mu}(r)$, which was known from the preceding experimental studies at Tien Shan \cite{ontienmuons3}:
$$
\varphi_{\mu}(r)=A\cdot {r^{-0.7}}
\exp\left(-\frac{r}{80}\right).
$$
Here, $A=5.95\cdot 10^{-4}$\,particles/m$^2$ is a normalization coefficient, and the distance to EAS center $r$ is expressed in meters. In former experiments at Tien Shan the function $\varphi_{\mu}$ was obtained by averaging the measurements of muon density which were made in EAS up to the distance $r \sim 200$\,m. With this function,
$$
N_{\mu}=\frac{\overline\rho_{undg}}
{\frac{1}{k}{\sum_{i=1}^{k}{\varphi_{\mu}}(r_{i})}},
$$
where $r_i$ is the  distance of the $i$-th group from the shower center.

\subsection{The underground neutron detector and its use for registration of cosmic ray muons}

\begin{figure*}
{\centering
\includegraphics[width=0.8\textwidth, trim=0mm 50mm 60mm 0mm]{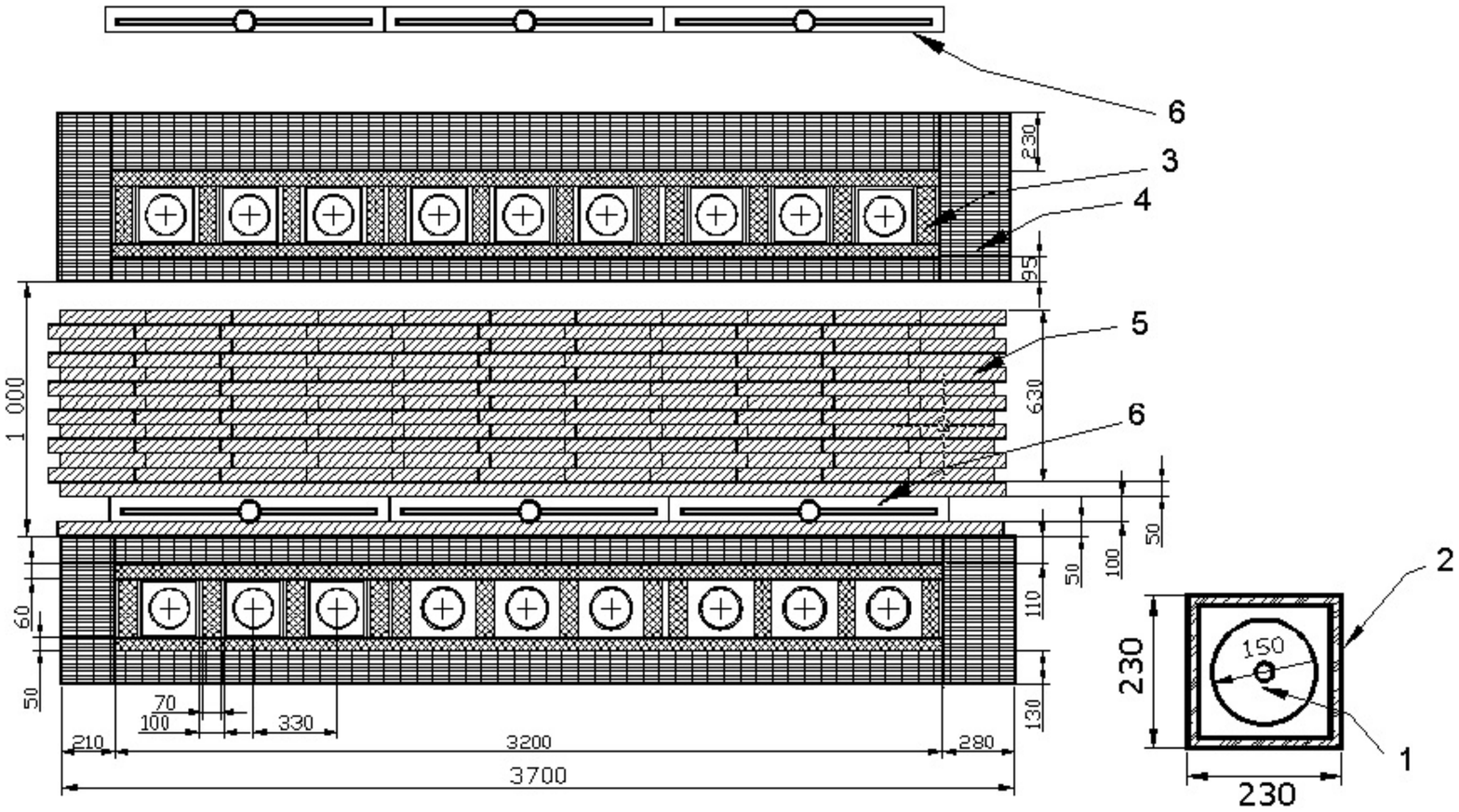}
\caption{Internal arrangement of the underground muon detector. \textit{1}---the gas discharge neutron counter, \textit{2}---light neutron moderator (polyethylene and wood), \textit{3}---lead target, \textit{4}---rubber, \textit{5}---iron absorber, \textit{6}---the coincidence telescope scintillators. The dimensions are shown in millimeters.}
\label{figiundgdete}}
\end{figure*}

An alternative way to study the muon component of cosmic rays at the Tien Shan mountain station is using an underground neutron detector which was incorporated into experimental complex at the modern stage of its development \cite{our2018_undg,ontien-nim2016,our2017_ontien}. The internal arrangement scheme of the underground detector is presented in  Fig.\,\ref{figiundgdete}. The detector is based on a set of neutron sensitive gas discharge counters surrounded by alternating layers of different materials: the heavy target absorber made of lead and iron, and the light neutron moderator of polyethylene, rubber, and wood. High energy cosmic ray muons penetrating into this detector give origin to evaporation neutrons, either through photonuclear excitation of absorber nuclei by the bremsstrahlung gamma rays emitted by muons, or by the channel of direct muon-nuclear interaction. Afterwards, the neutrons loose their energy down to thermal values in multiple collisions with light nuclei in the moderator layers which surround the counters. The counters can detect the presence of thermal neutrons due to the special filling which includes the BF$_3$ gas enriched by $^{10}$B isotope, such that the low-energy neutrons can cause the reaction $n$($^{10}$B, $^{7}$Li)$\alpha$ inside the counters.

The underground detector consists of a pair of separate, $(3.7\times 2)$\,m$^2$, units placed one above the other, as it is shown in Fig.\,\ref{figiundgdete}. Both units contain 9~neutron counters, each with geometrical sizes of $(\diameter 15\times 200)$\,cm$^2$. The whole detector set-up is hosted in the underground room of the Tien Shan station under a $\sim$$2000$\,g/cm$^{-2}$ thick rock layer which shields it from above to prevent penetration of ordinary hadrons, such as nucleons and pions, whose interactions could be an interfering source of evaporation neutrons. Thanks to this absorber, the attenuation factor of the hadron component of cosmic rays is of about $10^{-6}$. As well, the same absorber determines the minimum energy of muons capable to penetrate into the upper unit of underground detector as $\sim$5\,GeV. Due to an additional $\sim$$3500$\,g/cm$^{-2}$ thick mass of iron between the upper and lower detector units the latter has a somewhat higher energy threshold of muon registration of about $(10-12)$\,GeV (for vertical muon incidence).

Because of the said operation principle, an observable signal of muon passage through the underground detector is a series of electric pulses obtained from its neutron counters during a narrow time window, typically of a few milliseconds order. Further on such dense signal group will be referenced to as ``neutron event''. Sum count of detected pulses, named ``multiplicity'' of an event $M$, is evidently proportional to the total amount of produced evaporation neutrons, and the latter depends both on the number of muons which have simultaneously come through the detector, and on their energy.

By operation of the experimental data from the underground detector which will be presented further on, the length of the time window for calculation of the events multiplicity was selected as $4.5$\,ms after every EAS trigger. This was made with account of the average life time of evaporation neutrons in detector unit ($\sim$0.45\,ms) and of the intensity of background signals ($\sim$$2$\,s$^{-1}$) which would distort multiplicity measurement by randomly falling into the window.

\begin{figure*}
{\centering
\includegraphics[width=0.49\textwidth, trim=20mm 0mm 8mm 0mm]{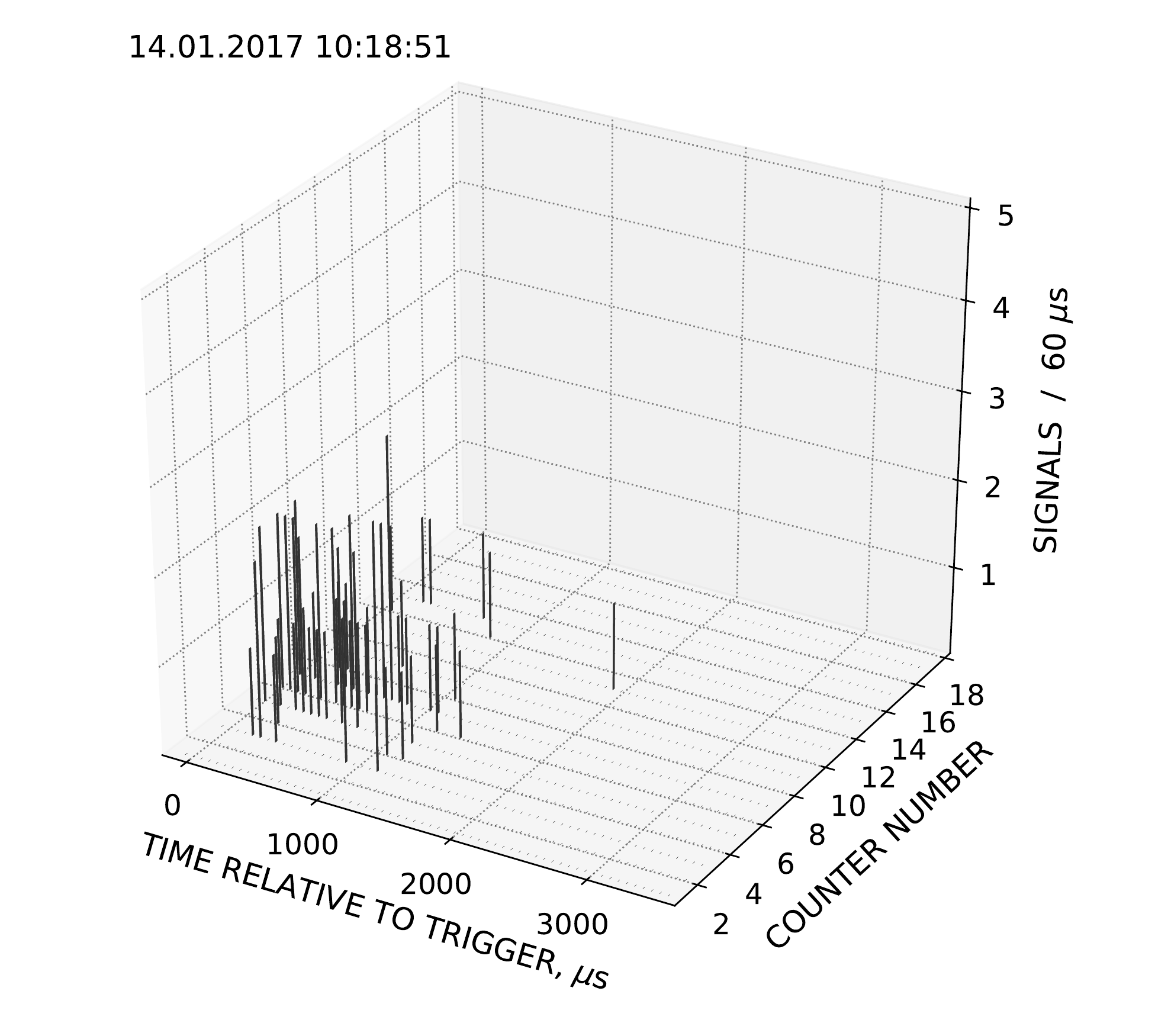}
\includegraphics[width=0.49\textwidth, trim=8mm 0mm 20mm 0mm]{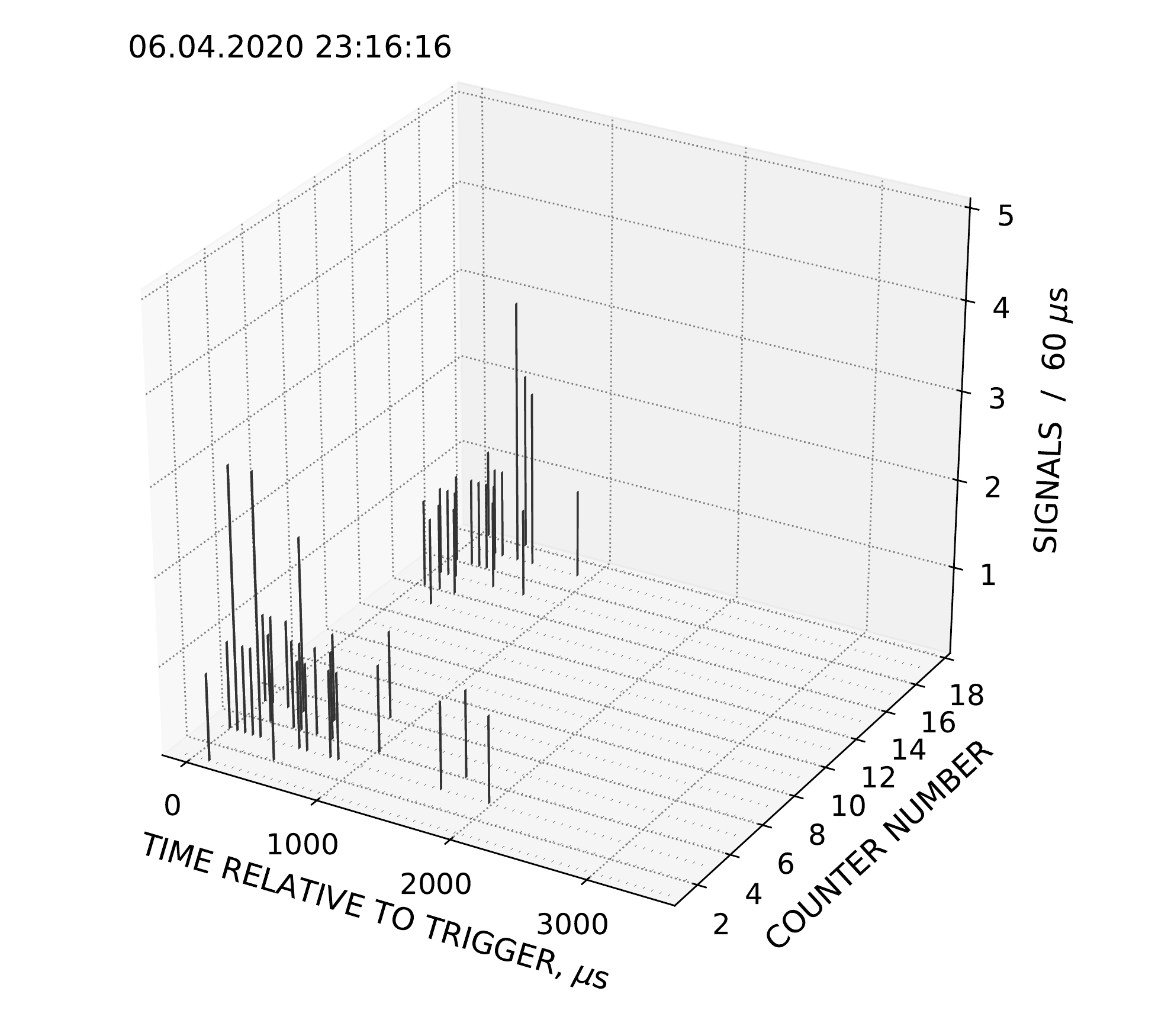}
\caption{Time distribution of neutron signals in the underground detector after EAS passage in the two sample events from Fig.\,\ref{figieassample}. The neutron counters \textit{1--9} correspond to the upper detector unit, \textit{10--18} correspond to the lower one. Zero point of time axes corresponds to the arrival moment of EAS triggers from the system of shower detectors.}
\label{figiundgsample}}
\end{figure*}

A sample of neutron response as it was seen in the underground detector by passage of the muons from a close extensive air shower is presented in Fig.\,\ref{figiundgsample}. In these two cases the multiplicities of the neutrons detected in the upper and lower units were, correspondingly, $63$ and $6$ in the left frame event, and $25$ and $31$ at the right. The primary energy of the EAS in these events was $1.2$\,PeV and $12$\,PeV.

\begin{figure}
{\centering
\includegraphics[width=0.47\textwidth, trim=0mm 0mm 0mm 0mm]{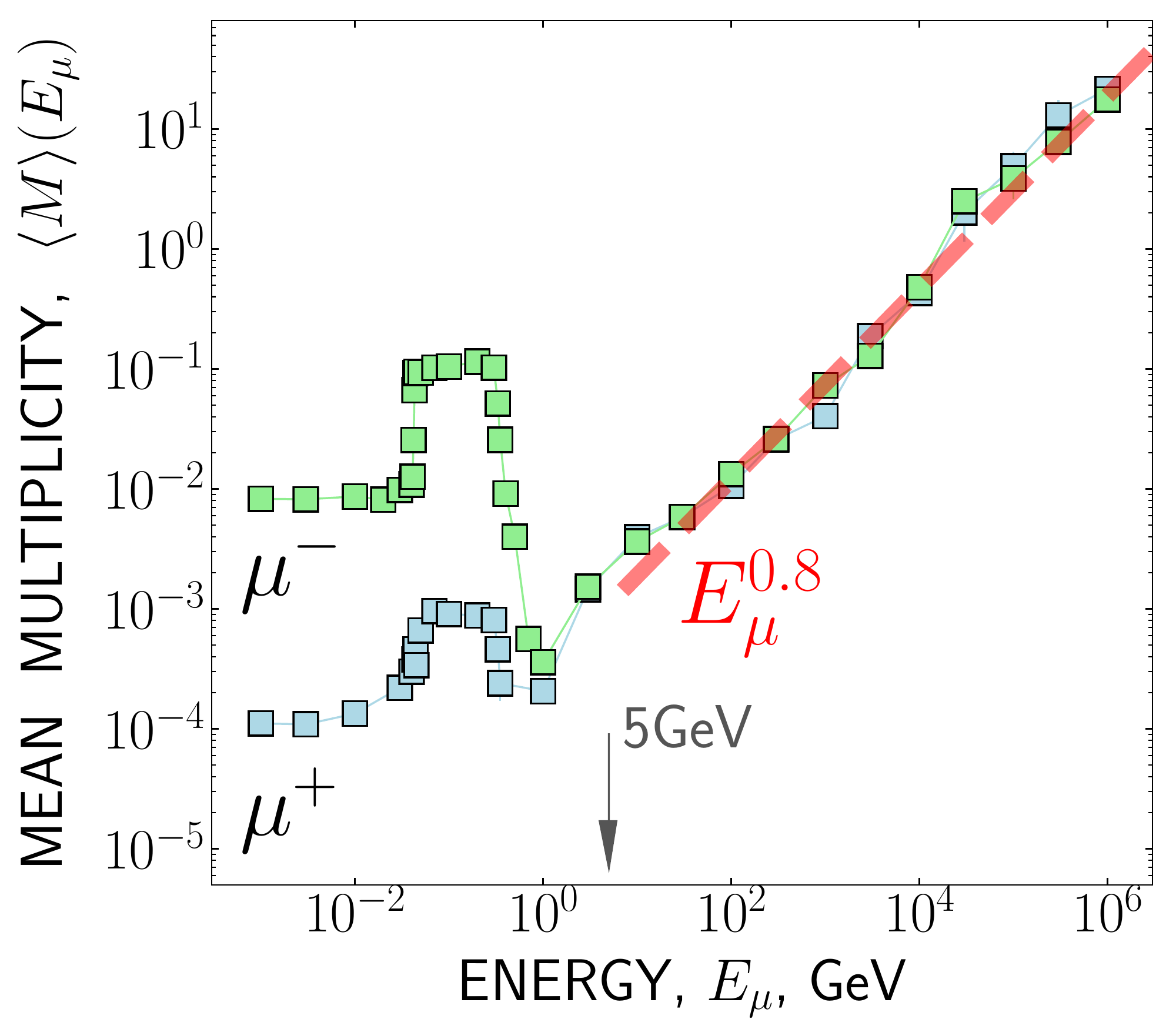}
\caption{Mean multiplicity $M$ of the detected neutron signals from the counters of an underground detector unit in dependence on the energy of interacting muon $E_{\mu}$ (the result of a Geant4 simulation). The straight dashed line corresponds to the power type function, $M\sim E_\mu^{0.8}$. The lower energy threshold of muon registration in the underground detector is marked by a vertical arrow.}
\label{figiundgcalibri}}
\end{figure}

The process of muon induced neutron generation in the underground detector was studied in detail by a Geant4 simulation of muon interactions which was made for the specific configuration model of the detector. The key result of this simulation is presented in Fig.\,\ref{figiundgcalibri} as distribution of the anticipated mean multiplicity of neutron event $M$ in dependence on the energy $E_\mu$ of a single muon hitting the detector, $M(E_\mu)$. In the range of $E_\mu$ values essentially below 1\,GeV this dependence has a bump which corresponds to reactions of muon capture by the atomic nuclei in absorber. In the practically interesting  operation range of the underground detector which starts only from 5\,GeV, and is strictly limited from below by penetration possibility of muons through the overlying soil, the dependence of neutron multiplicity on muon energy is of a simple power type, $M\sim E_\mu^{0.8}$.

The threshold of assured registration of a single muon in the underground detector is defined by the minimum value of muon energy $E_\mu$ at which the average count of neutron signals occurs above zero, $M\geqslant 1$. As it follows from Fig.\,\ref{figiundgcalibri}, for the considered detector set-up this lower  limit lays around the values of $E_\mu\simeq (20-30)$\,TeV. Thus, in contrast to the muon hodoscope of the \HADRON experiment the underground neutron detector is an instrument for preferable study of a rather high-energy muon component. Besides, the available information on neutron multiplicities principally opens an opportunity for estimation of the muon energy in each particular event.

\section{XREC families and muon quantities in EAS}
\label{sechadron}

\subsection{Experimental data}

During the whole operation time of \HADRON experiment it was found 1665 families in the upper $G$-block of the Tien Shan XREC which it succeeded to bind with corresponding EAS. Another 991 families were found in the lower $H$-block.

\begin{figure*}
{\centering
\includegraphics[width=0.5\textwidth, trim=0mm 0mm 0mm 0mm]{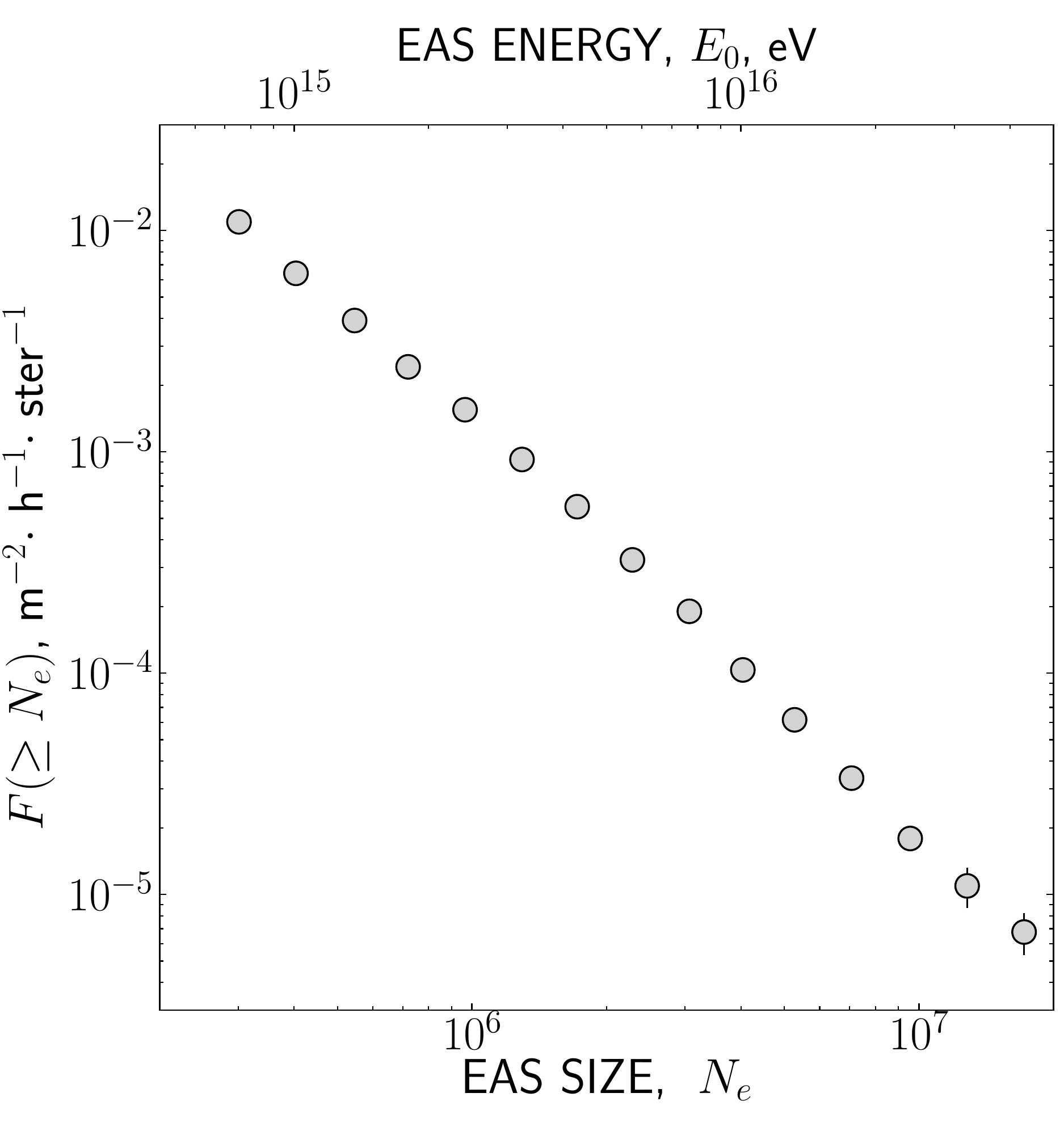}
\caption{The integral spectrum of shower sizes in EAS events detected in the \HADRON experiment.}
\label{figihadrospc}}
\end{figure*}

For better unambiguity of the EAS---XREC events correlation the search for family candidates for each shower was made between the families which comply the two, rather liberal, criteria: $N_\gamma\geqslant 1$ and $\sum{E_\gamma}\geqslant 10$\,TeV, where $N_\gamma$ is the total amount of electron-photon cascades detected in a family, and $\sum{E_\gamma}$ is the sum energy of the cascades. As well, only the showers with the size $N_e$ above $10^5$ were taken into account at this search, since for smaller EAS the resulting statistical distributions might be distorted because of particulars of the shower trigger elaboration algorithm accepted in the \HADRON experiment. The total statistics of the EAS events included into analysis amounts to $33172$. As it is seen in Fig.\,\ref{figihadrospc}, the histogram of these events over the size parameter is of a characteristic power type in the whole range of considered $N_e$ variation, and does not have any  distortion which could be caused by threshold effects at the side of small EAS. The regular form of the spectrum means correct operation of the algorithm of showers selection, without implicit preference of any specific kind of events (\eg by the type of primary particle).

\begin{figure*}
{\centering
\includegraphics[width=0.5\textwidth, trim=0mm 0mm 0mm 0mm]{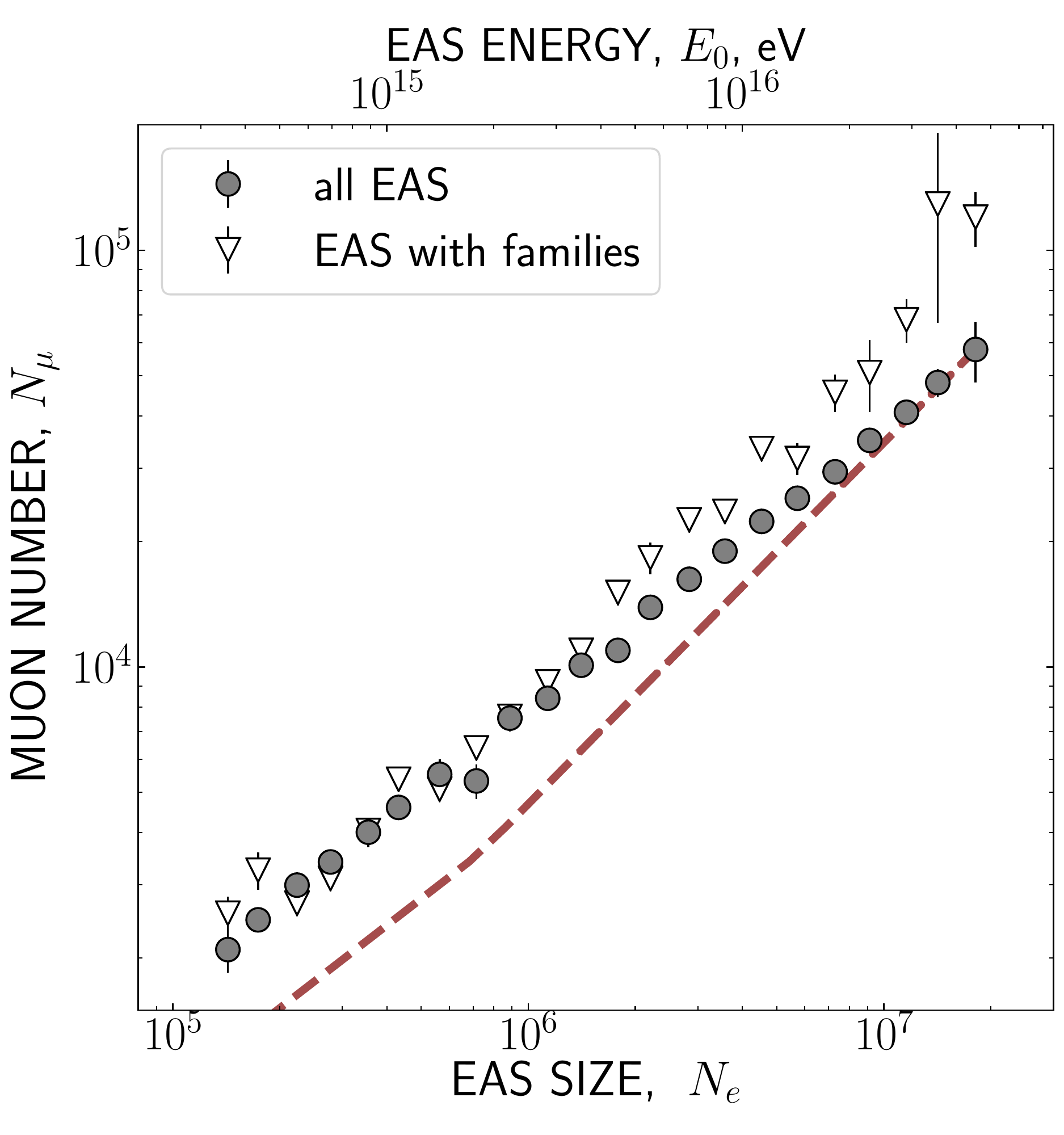}
\caption{The average number of muons in EAS, $N_\mu$, in dependence on the mean size $N_e$ and primary energy $E_0$ of the shower. The dashed line indicates the anticipated $N_\mu(N_e)$ distribution in supposition that all family events were caused by the showers originating from primary protons (see Section\,\ref{sectiaccu} for discussion).}
\label{figixrec}}
\end{figure*}

Later on, it was revealed that interesting information on peculiar properties of hadron interaction at high energies may be gained from the multiplicity data on muon component in the showers which it succeeded to combine with their corresponding counterparts in X-ray films. In Fig.\,\ref{figixrec} it is presented a correlation dependence $N_\mu(N_e)$ between the average amount of muons $N_\mu$ in the showers and their average size $N_e$. In this plot the experimental $N_\mu(N_e)$ distribution is shown twice: both for the whole statistics of the registered EAS events without any specific selection (shown with solid circles), and only for those showers which were accompanied by a gamma- or hadron family in either of two blocks of the Tien Shan XREC (shown with open triangles). As it is seen in the figure, the $N_\mu(N_e)$ distributions of both types superimpose over each other in the region of small EAS sizes, but starting from the point of $N_e\approx 10^6$ the muon content in the showers with families does systematically exceed the common level. At the end of the investigated range of shower sizes, $N_e\sim 10^7$, the relative difference between the two experimental distributions in Fig.\,\ref{figixrec} is nearly twice as large as the mean $N_\mu$ value calculated over the whole EAS dataset.

\subsection{The anomaly of muon content in the showers with families}
\label{sectimuoco}

As it follows from numerous simulations, \eg \cite{2007aragats_simul_like_tien,weber_icrc1999,2005kascade}, generally an enhanced muon abundance in any specific group of extensive air showers means a corresponding excess of the average atomic mass of cosmic ray nuclei these EAS originate from. For example, in \cite{2007aragats_simul_like_tien} the simulation is presented which was made for the \GAMMA installation at Aragats, constructively analogous and situated at the same altitude above the sea level as the Tien Shan one. According to their data, the average number of muons in a shower originating from iron nucleus must be of up to $(30-40)$\% above that of a proton-induced one. Therefore, an excess of the average muon content in the showers specially selected by the presence of families over the whole statistics of EAS events in Fig.\,\ref{figixrec} should be interpreted as predominance of heavy nuclei among the cosmic ray particles which were progenitors of the EAS with families.

However, multiple results of simulation studies of the process of family production in XREC, such as \cite{tamada_simul_1994_e0_from_ne,onxrec_mass_2009,onxrec_mass_2017,onxrec_mass_2018}, are quite opposite,  though intuitively reasonable: since the maximum of shower development in EAS which stem from the light particles (p and He) is located essentially deeper in the atmosphere than for heavy nuclei (Fe), it is the light component of primary cosmic rays which is most probable to carry the largest part of its initial energy with minimum losses up to the level of the Tien Shan station, and to generate XREC families there with highest efficiency. Thus, according to \cite{tamada_simul_1994_e0_from_ne}, about 80\% of family events detected at the Tien Shan station must be originated from interaction of cosmic ray protons and helium nuclei.
Lighter composition of cosmic ray primaries in the events with families was confirmed also experimentally, by an analysis of specific properties of the showers which had accompanied such events in the \HADRON experiment \cite{shaulov_composition_2003}.
On the other hand, if this were true the mean number of muons $N_\mu$ in the shower events selected by the presence of families should be essentially lower, but not higher, then in general.

Therefore, direct contradiction exists between the two conclusions on average mass composition of the cosmic ray particles which could be preferred source of families in the Tien Shan XREC: according to the EAS data these must be protons and helium nuclei, while the muon content in such events rather corresponds to the heavy, Fe-like primaries. This contradiction means an inadequacy of the particles interaction models our interpretation of family production is based on to the processes which actually take place at collision of high energy cosmic ray particles with air nuclei.

It should be also noted that presently the information on composition of primary cosmic rays in the $(1-100)$\,PeV energy range is rather uncertain: in contrast to supposition on predominance of light primaries which was necessary in \cite{shaulov_composition_2003} to explain the intensity of gamma-families at the \HADRON installation, heavy nuclei (Fe) should prevail at a few tens of PeV according to the measurement data and simulations of the \KASCADE group \cite{2005kascade,kascade2013}. In some other works, such as \cite{2007aragats_simul_like_tien} or \cite{horandel2016}, the lighter composition was still claimed for the same energy range.

As it was just mentioned above, the border point at $N_e\approx 10^6$, where the divergence starts to be observable between the two $N_\mu(N_e)$ distributions in Fig.\,\ref{figixrec}, corresponds to the extensive air showers with primary energy $E_0\approx 3$\,PeV, \idest to the events which belong to the knee region of the cosmic ray spectrum. Thus, the observed difficulty with explanation of mass composition of the cosmic ray primaries which were preferred ancestors of the showers with families reveals itself just at the knee of the spectrum.

It should be kept in mind that the energy threshold of muon detection by the hodoscope of the \HADRON experiment was of about $5$\,GeV, as defined by the thickness of soil absorber above the underground room of the Tien Shan station. Consequently, because of steeply diminishing spectrum of cosmic ray muons all the results on the muonic EAS component gained at those times and discussed here relate mostly to the muons with comparatively moderate energies of a few GeV order. As well, the selection of EAS events by the presence of gamma-hadron families in X-ray films is automatically equivalent to considering only those showers whose cores did hit the Tien Shan XREC and the muon hodoscope beneath, since the families emerge from interaction of the most energetic EAS hadrons which concentrate in a narrow space around the shower axis. For this reason the conclusion on observation of a peculiarly large muon content in EAS with families and of its discrepancy with simulation models relates specifically to the central region of EAS. With account to the average diameter of gamma-hadron families the typical lateral size of this region can be estimated as $(5-10)$\,cm only, \idest it remains essentially below a meter.

\subsection{On estimation accuracy of shower parameters}
\label{sectiaccu}

The result on excess of muon content in the showers with families obtained at the \HADRON installation was rather unexpected, so additional analysis of possible errors in operation of experimental data is required.

The discrepancy between the two distribution types in Fig.\,\ref{figixrec} might erroneously arise from systematic deviation of either the shower size or muon number estimates in the EAS specifically selected by the presence of families. Since the preferred source of such events were the light cosmic ray nuclei, as it follows \eg from  \cite{tamada_simul_1994_e0_from_ne} and \cite{shaulov_composition_2003}, their actual muon numbers should be systematically less than in average. If to suppose also that the general shape of $N_\mu(N_e)$ dependency in the EAS with families was correctly defined in the experiment, and it was only the absolute value of muon number which was wrongly overestimated, the proper position of that dependency should be as it is shown by the dashed line in Fig.\,\ref{figixrec}. The latter was obtained as interpolation of the points marked there with triangles, and displaced in direction of smaller $N_\mu$ values by such a way that the points at its tail were superimposed over the points of the all-EAS distribution. As it follows from the plot, the difference between the triangular marks of experimental points and the dashed line would disappear if to multiply by $(2.5-3)$ the $N_e$ parameter for the EAS with families, or to decrease twice their $N_\mu$ values.

As explained in Section\,\ref{sectishw}, the shower size $N_e$ in the considered experiment was set independently of any apriori model supposition, as an integral of lateral distribution of particles density, experimentally measured and approximated by an NKG type function. The accuracy of such $N_e$ estimates depends only on precision of the approximation, and the latter is primarily determined by exactness of the EAS axis position. In the central part of the \HADRON installation the axis was located by the signal distribution in ionization chambers with accuracy $\pm$$0.25$\,m, which is $(3-10)$ times better than in the case if the data on EAS were only accessible from the system of shower particles detectors. An analysis of different approximation techniques applied for the task, such as the least square method and the method of maximum likelihood, has shown that all of them give close results with  relative error of $N_e$ estimation of about $(5-10)$\%, which remains independent of the type of primary cosmic ray particle. It could be hardly expected in such a case to meet systematically the errors higher by an order of magnitude which would be necessary to explain the deviation of the points marked with triangles from the assumed dashed line in the plot of Fig.\,\ref{figixrec}, so the supposition on large incorrectness of $N_e$ estimates should be omitted.

\begin{figure*}
{\centering
\includegraphics[width=0.37\textwidth, trim=0mm 7mm 8mm 0mm]{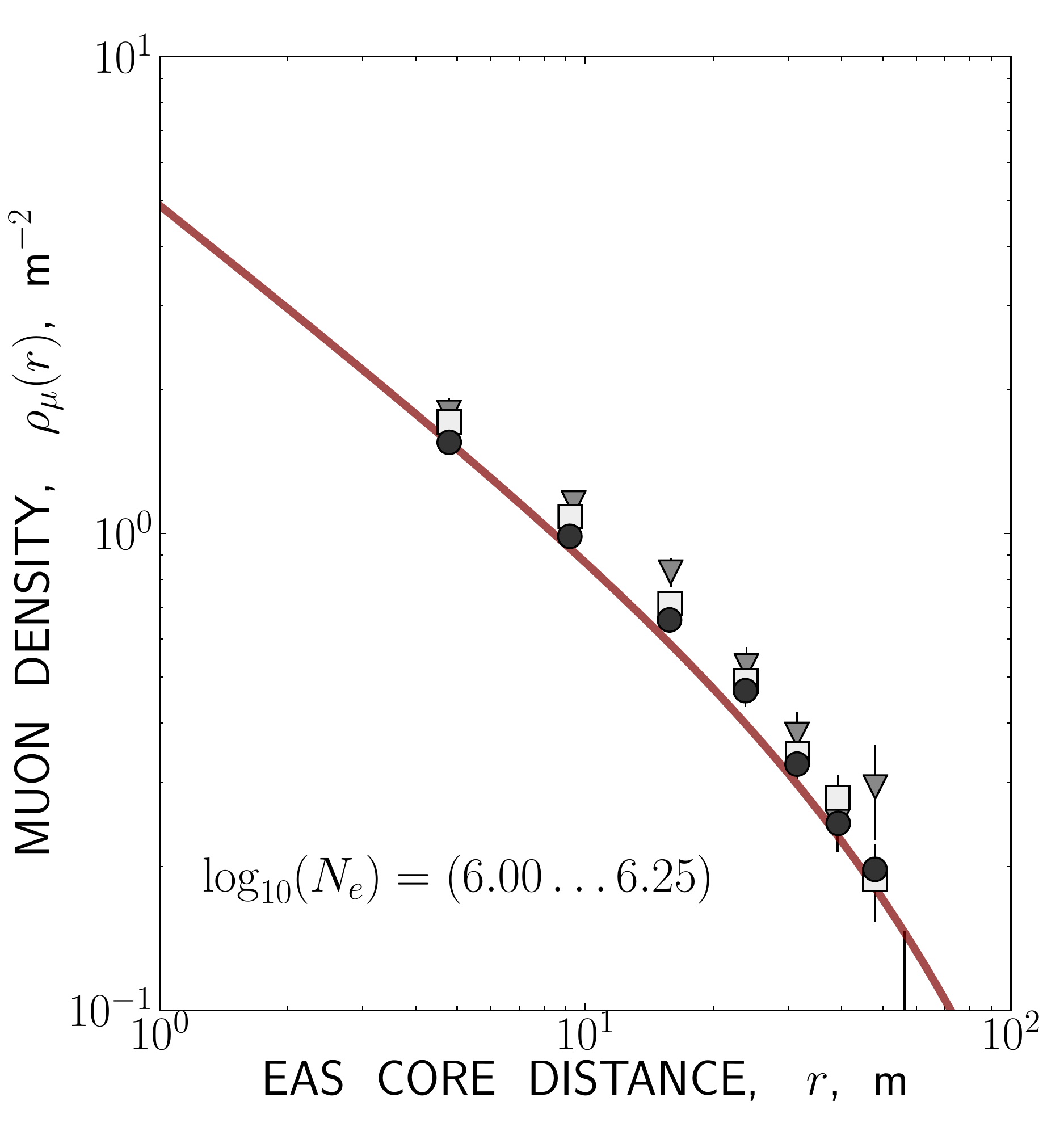}
\includegraphics[width=0.37\textwidth, trim=0mm 7mm 8mm 0mm]{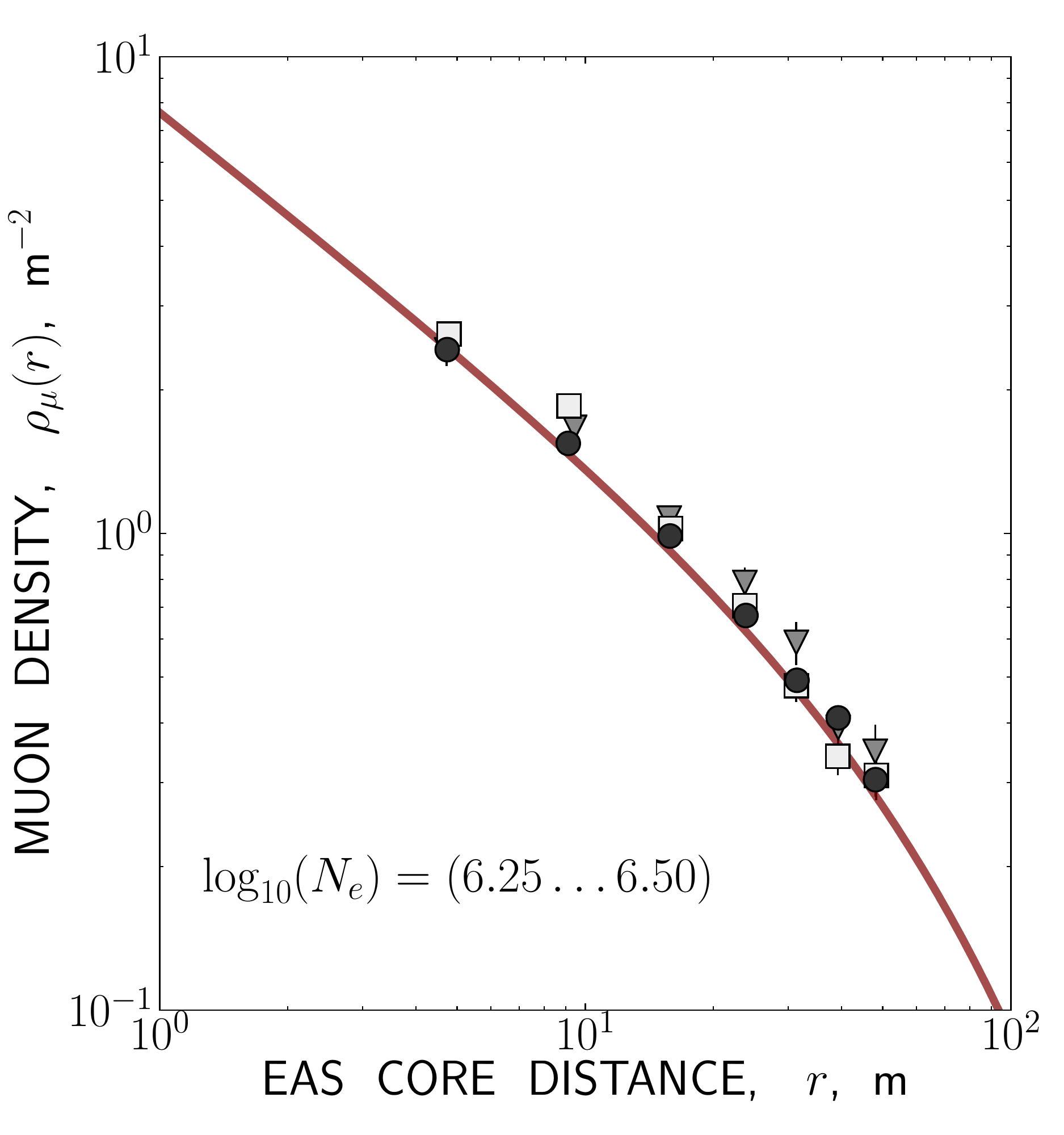}
\\
\includegraphics[width=0.37\textwidth, trim=0mm 7mm 8mm 0mm]{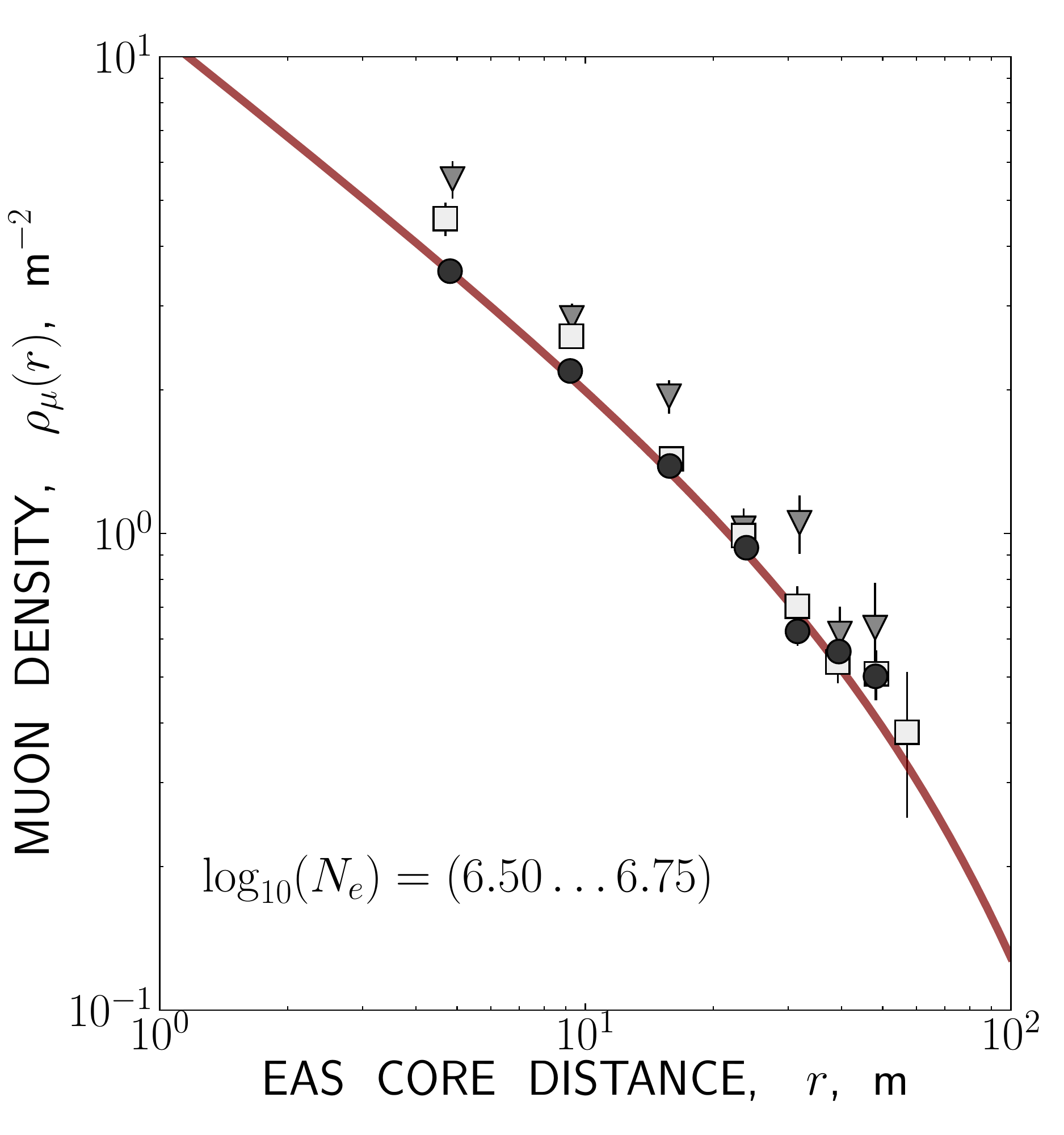}
\includegraphics[width=0.37\textwidth, trim=0mm 7mm 8mm 0mm]{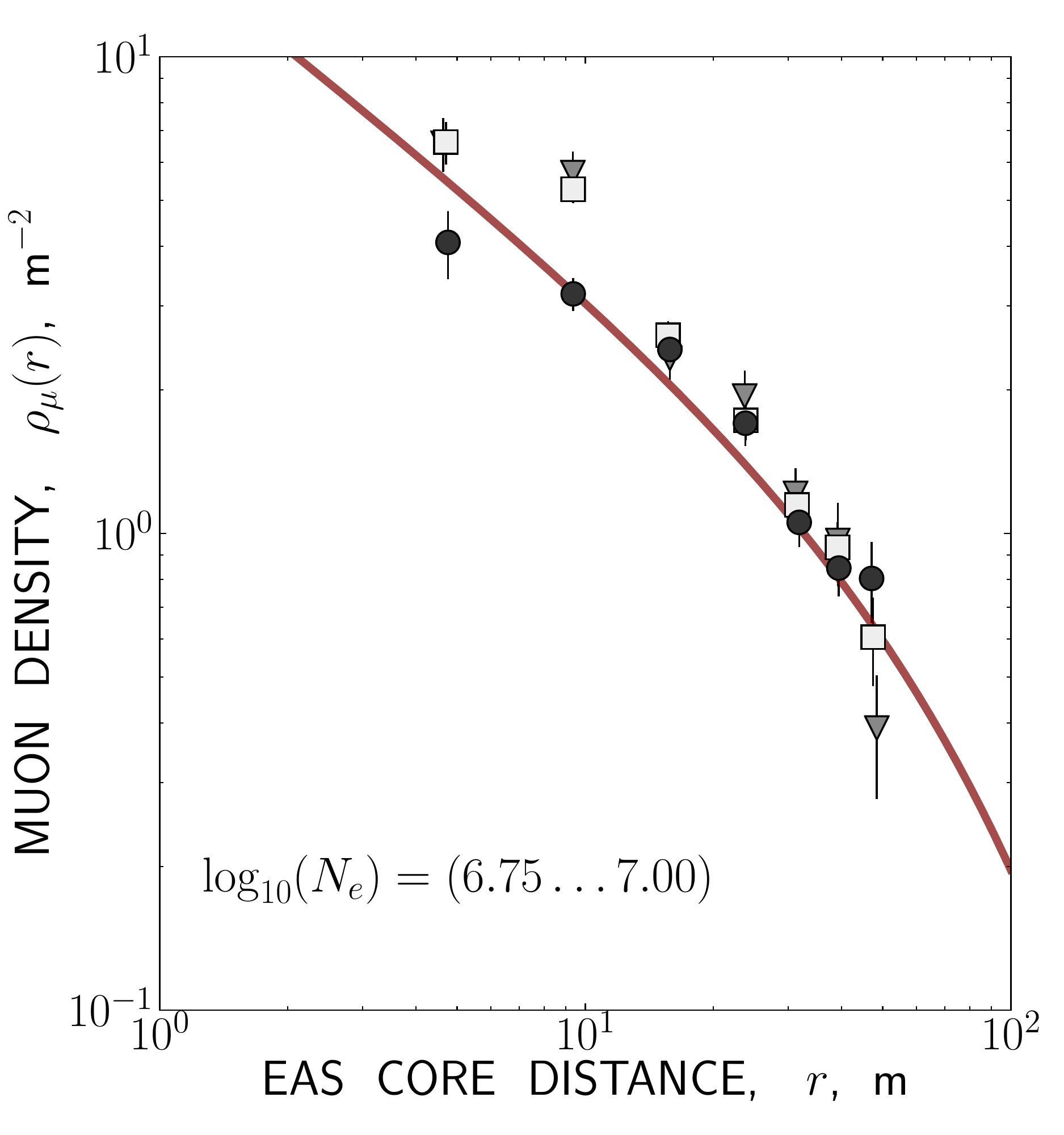}
\\
\includegraphics[width=0.37\textwidth, trim=0mm 10mm 8mm 0mm]{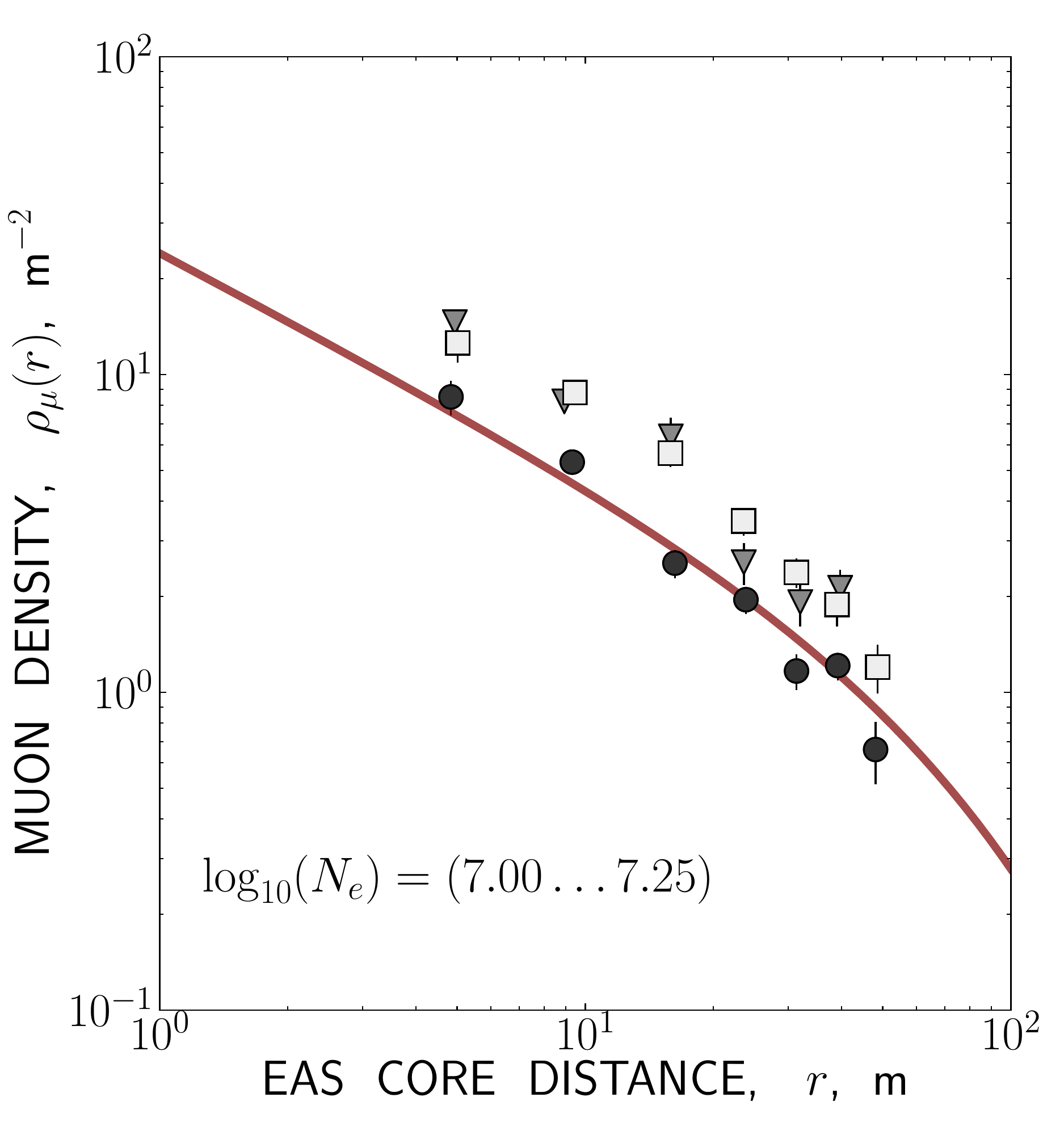}
\includegraphics[width=0.37\textwidth, trim=0mm 10mm 8mm 0mm]{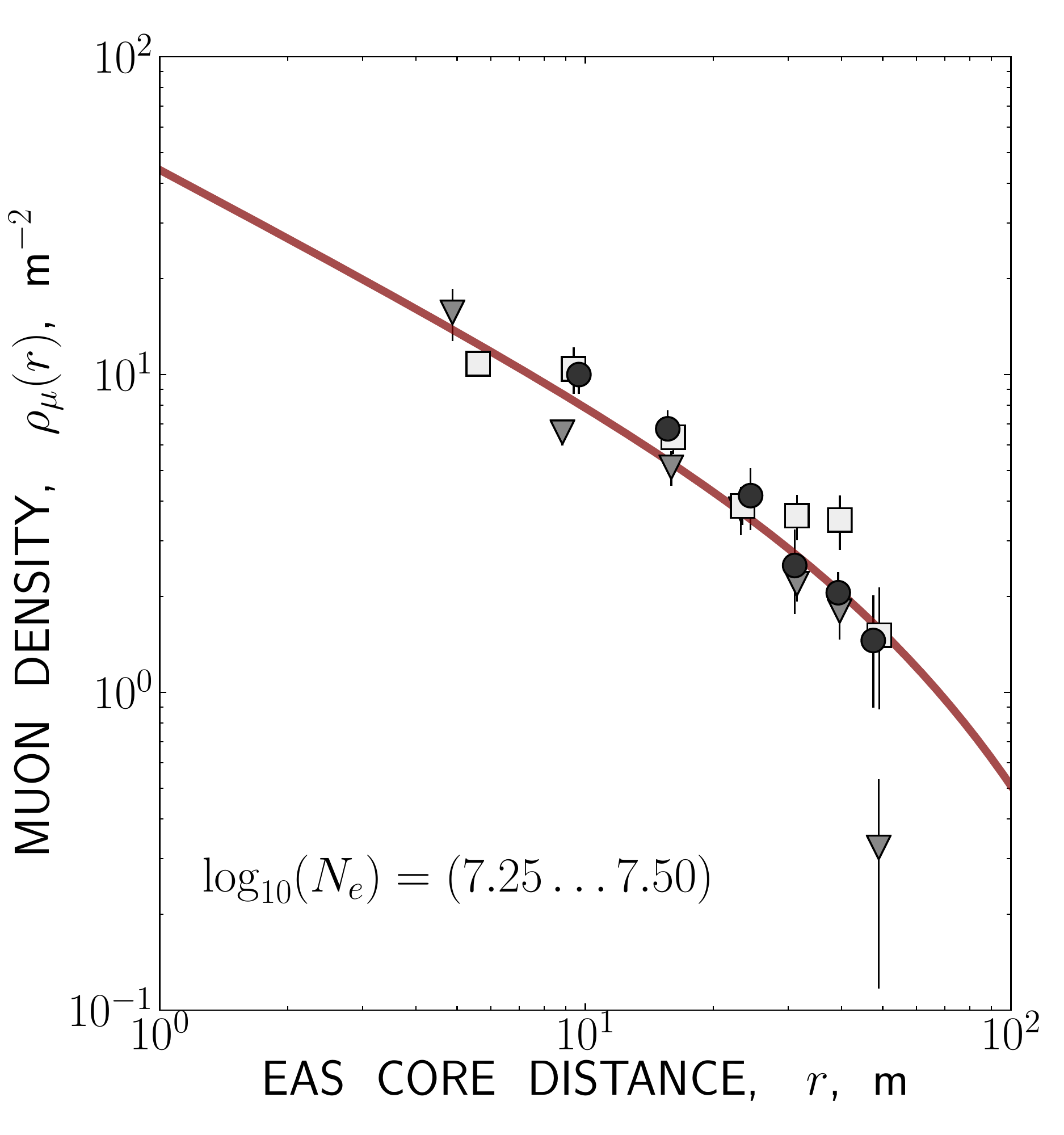}
\caption{The average lateral distribution of muon density in all EAS events (circles), and only in the events with gamma- (squares) and hadron (triangles) families detected in the upper and lower blocks of the Tien Shan XREC. The smooth continuous curves correspond to the universal distribution function $\varphi_{\mu}(r)$ mentioned in Section\,\ref{secmuohodo}.}
\label{figrhomu}}
\end{figure*}

\begin{figure*}
{\centering
\includegraphics[width=0.5\textwidth, trim=0mm 0mm 0mm 20mm]{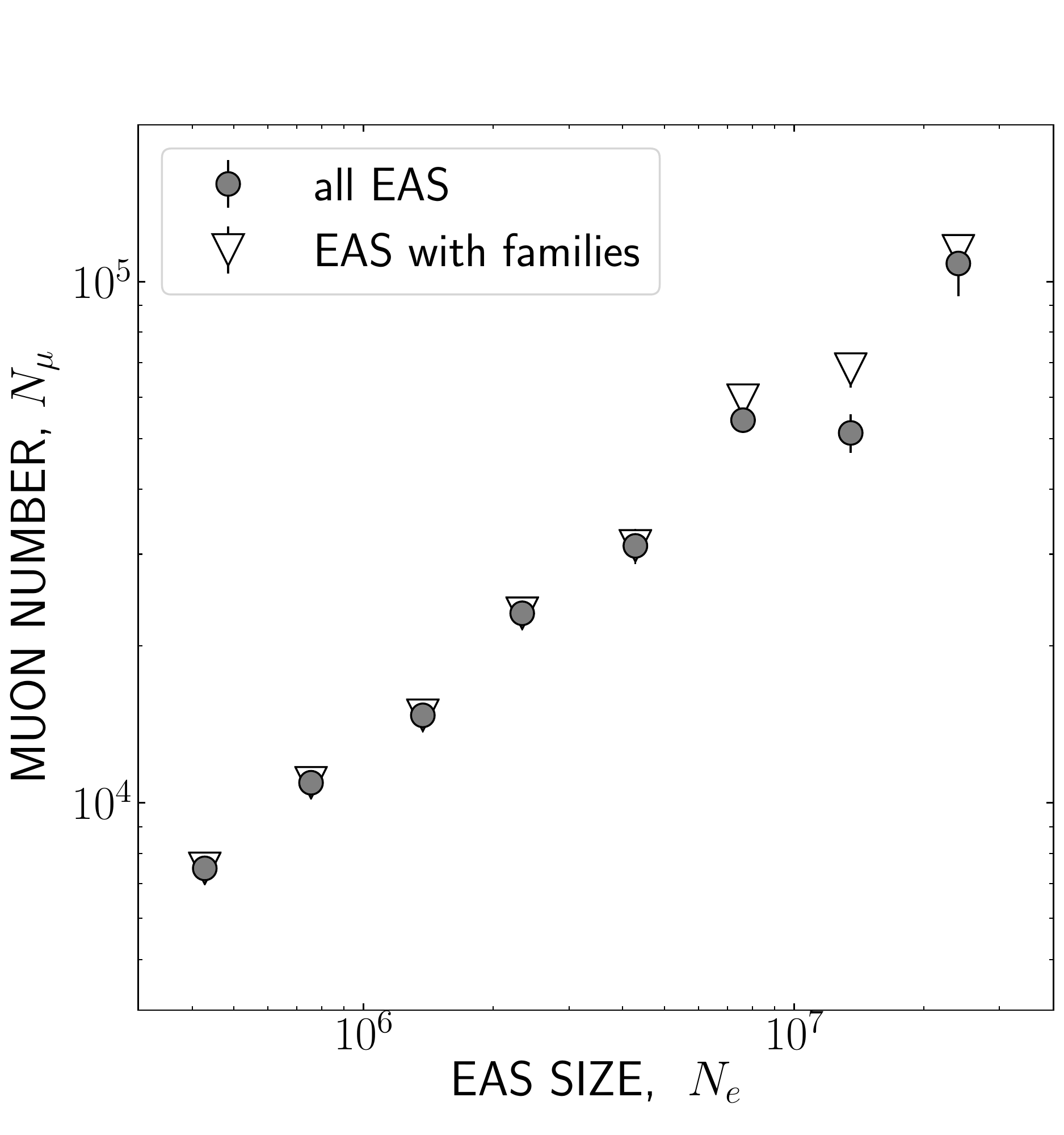}
\caption{The dependence between the average $N_\mu$ and $N_e$ values for the $N_\mu$ estimates calculated using the individual distribution curves of muon density in the EAS (see text). Designations are the same as in Fig.\,\ref{figixrec}.}
\label{figinmu_vs_ne2}}
\end{figure*}

As for the determination accuracy of muon number in EAS, it should be primarily noted that at the time of \HADRON experiment the method of $N_\mu$ estimation by hodoscope data, as well as the hodoscope device itself, were remaining  the same as during the previous period of EAS investigation at Tien Shan, when the whole technique necessary for the purpose was designed and described  in detail in \cite{ontienmuons3}.  As explained in Section\,\ref{secmuohodo}, this method is based on using an universal distribution function $\varphi_{\mu}$ which had been deduced from averaging of experimental data. The total number of muons in a shower $N_\mu$ was calculated as a quotient of division of the muon density measured by the hodoscope to the standard $\varphi_{\mu}$ distribution. Then, a systematic error of $N_\mu$ estimates in the showers with families might be caused by deviation of the actual distribution of muon density in those EAS events from the function $\varphi_{\mu}$. 

To check such opportunity a comparison was made between the lateral distribution of muon density $\rho_\mu$, as it was seen in all EAS, and distributions calculated only for the showers in which the families were detected either in the $G$- or $H$-block of the Tien Shan XREC (correspondingly, $\rho_\mu^\gamma$ and $\rho_\mu^h$). Such distributions are presented in Fig.\,\ref{figrhomu}, separately for several groups of showers with the close sizes $N_e$, and together with a normalized distribution of the universal $\varphi_\mu$ type in each case. As it follows from the figure, the shape of both the $\rho_\mu$ and $\rho_\mu^{\gamma,h}$ distributions, as they were measured in the \HADRON experiment, generally agree with the function $\varphi_\mu$.

The method described in \cite{ontienmuons3} is based on the measurements of muon density which were actually made in a limited interval of distances from the shower axis, $r\leqslant 60$\,m, as a consequence of specific configuration of the Tien Shan muon detector. (There were two hodoscope parts there: the one placed in the central underground room, and another in the adjoining linear tunnel with the length of $\sim$$50$\,m; see Fig.~\ref{figihadroundg}). Within these limits only a $\sim$$40$\% share of all EAS muons can be directly detected, so the result of $N_\mu$ calculation might be biased if the actual distribution of muon density at the uncontrolled EAS periphery would deviate essentially from the function $\varphi_\mu$.
Nevertheless, diminishing twice the $N_\mu$ values in EAS with families which is necessary to agree them with position of the dashed line in Fig.\,\ref{figixrec} demands to suppose that the distribution of EAS muons practically terminates already at distance $r \simeq 100$\,m. Such alternative seems to be too radical variant.

The visible excess in muon numbers could be reduced also if to suppose that it is the far shower periphery at $r\gtrsim 100$\,m only where the lateral distribution of muon density in the events with families goes systematically more steeply than in general, but at distances $r\leqslant 100$\,m it remains in agreement with the universal function $\varphi_\mu$.
This possibility was verified by considering in each registered EAS event the individual curves of muon density $\rho_\mu(r_i)$, $i=1\ldots n$, which were immediately measured in $n$ hodoscope groups around the shower center, $r_i\leqslant 60$\,m, and extrapolating them to larger distances with the functions of the same type as the distribution $\varphi_\mu$, \idest a power function modulated by an exponent,  $\phi_\mu(r)=a\cdot r^{-b}\cdot \exp(-r/c)$. In each particular EAS event the $a$, $b$, and $c$ parameters were defined from the equation of maximum likelihood, using the experimental measurements of muon density $\rho_\mu$:
$$
\sum\limits_{i=0}^n\left(\rho(r_i)-\phi_\mu(r_i)\right)\left(\frac{1}{a}-\ln r_i +\frac{r_i}{c^2}\right)=0.
$$
With such extrapolation, the total amount of muons in an EAS can be calculated as an integral
$$
N_{\mu}=\int\limits_0^\infty \phi_\mu(r)\cdot rdr=2\pi ac^{2-b}\Gamma(2-b),
$$
where $\Gamma(.)$ is gamma-function. In Fig.\,\ref{figinmu_vs_ne2} the correlation is presented between the average numbers of electrons and muons in EAS, analogous to what is shown in Fig.\,\ref{figixrec} but for the $N_\mu$ estimates which were calculated in the way described here. As it follows from Fig.\,\ref{figinmu_vs_ne2}, taking into account the individual distribution curves of muon density in EAS, instead of the universal $\varphi_\mu$, permits to decrease essentially the visible overabundance of muon content in the EAS with families, though not completely when it comes to the showers of largest sizes, those with $N_e\sim 10^7$.

Thus, the interpretation problem of a rather odd result on muon multiplicities in EAS obtained in the \HADRON experiment is connected with absence at that time of the direct measurement data on behavior of muon density $\rho_\mu(r)$ at EAS periphery. If to suppose that in the showers with families the lateral distribution of muons at distances $r\gtrsim 100$\,m still remains in agreement with the universal function $\varphi_\mu$, then a significant excess of muon numbers $N_\mu$ in such events follows from the $\rho_\mu(r)$  measurements made closely to EAS center, at $r\leqslant 60$\,m. This is the situation reflected in Fig.\,\ref{figixrec}, and the contradictory conclusions on composition of primary cosmic ray particles drawn from this figure were discussed in Section\,\ref{sectimuoco}.

In contrary, in the case if the muon distribution actually breaks down to zero at distances $r\geqslant 100$\,m, the origin of EAS events with families yet can be explained as being preferably connected with the light cosmic ray nuclei. This is also an important conclusion, since it indicates to predominance of protons in the knee region of the cosmic ray spectrum which does contradict, again, to some contemporary views on the subject mentioned in Section\,\ref{sectimuoco}.

Finally, it should be noted that according to the recent simulations \cite{muoldfatperiphi} for the proton and iron type primaries the general shape of the lateral distribution of muon density in the $(10-3000)$\,PeV EAS does not depend on the kind of primary particle up to $(10^2-10^3)$\,m distance from the shower axis. No such dependence was found also among the measurement results and simulation data of the \KASCADE experiment \cite{muoldfatperiphi_kascade} which relate to detection of EAS particles at the sea level. Thus, the first of the two alternatives suggested above for explanation of anomalous  muon content in the EAS with families seems to be more likely, \idest it is the interpretation of experimental data reflected in the plot of Fig.\,\ref{figixrec}, not in Fig.\,\ref{figinmu_vs_ne2}, which corresponds more adequately to reality.

\section{The underground neutron events and the muon component of EAS}
\label{secundg}

As discussed in Instrumentation section, the next exploration stage of cosmic ray muons at the Tien Shan mountain station was connected with the underground detector where the signal of evaporation neutrons was used as indicator of muon passage. Since the beginning of the year~2015 this detector was operating together with the detector system of the EAS electron component, and registration of neutron signals there was synchronized with the common trigger from the shower installation. Thus, investigation of the muon component of EAS initiated at Tien Shan by the \HADRON experiment was continued further on using quite another kind of experimental technique.

\begin{figure*}
{\centering
\includegraphics[width=0.49\textwidth, trim=0mm 0mm 0mm 0mm]{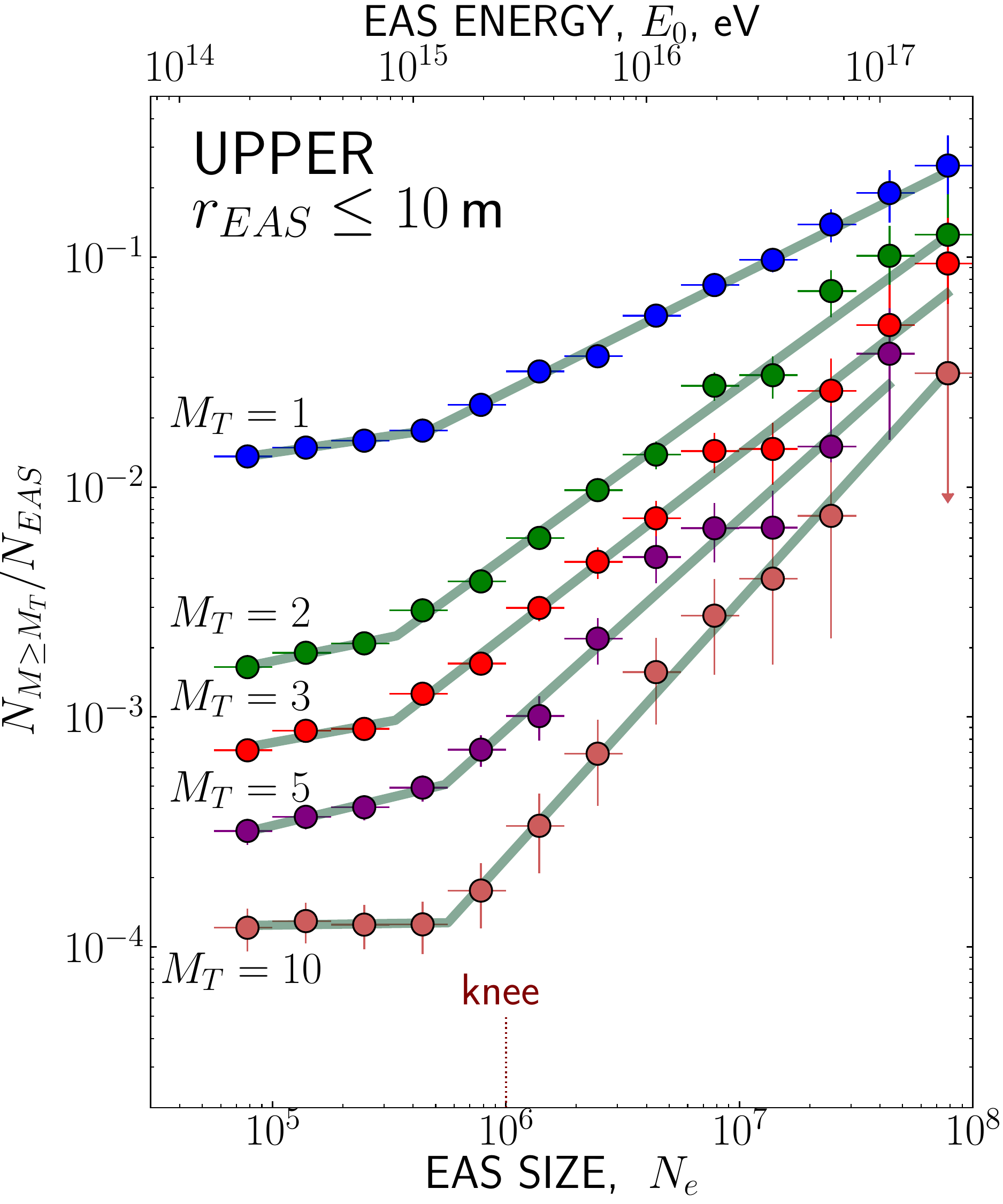}
\includegraphics[width=0.49\textwidth, trim=0mm 0mm 0mm 0mm]{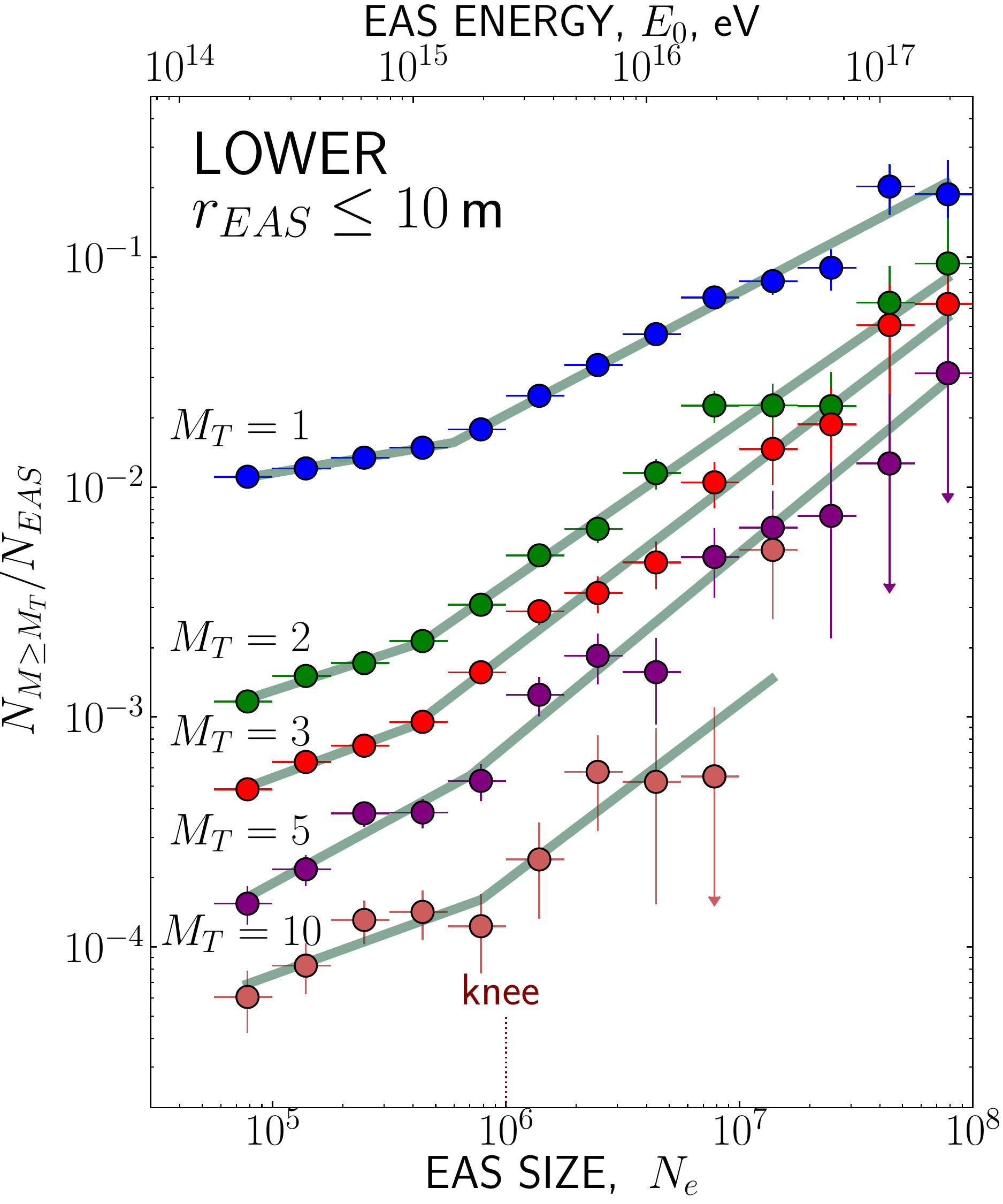}
\caption{The average share of the EAS
which have generated the neutron events with multiplicity \mbox{$M\geqslant M_T$} in the underground detector, in dependence on the mean shower size $N_e$, and on the minimum threshold multiplicity $M_T$ of neutron signals in an event. The counts of neutron events $N_{M\geqslant M_T}$ are normalized to the total amount $N_{EAS}$ of showers with the axis distance $r_{EAS}$  below 10\,m from the center of the underground detector. The thick broken lines correspond to the best fit approximation of experimental points with a piecewise-power function (see text).}
\label{figiundgevecounts}}
\end{figure*}

The principal result newly obtained during the last research period of EAS muons at the Tien Shan station is shown in Fig.\,\ref{figiundgevecounts} as distribution of the relative share of extensive air showers which were accompanied by an underground neutron event from muon interaction, in dependence on the average size of EAS $N_e$, and with account to the minimum threshold multiplicity $M_T$ of neutrons detected in an event. The latter parameter played its limiting role by calculation of the statistics of neutron events for the presented curves: only the cases with multiplicity $M\geqslant M_T$ where taken into account for each corresponding distribution. After calculation of events counts the resulting number of the showers with neutron events $N_{M\geqslant M_T}$ was normalized to the total amount $N_{EAS}$ of all showers which belong to the given range of EAS sizes $N_e$. In every registered EAS case both units of the underground neutron detector were considered independently of each other, so there are two similar plot panels in Fig.\,\ref{figiundgevecounts} which relate separately to the upper and lower detector unit.

To reduce the influence of the random muon background and of possible statistics distortion by the particles of peripheral EAS, only those showers were included into the events counts for  Fig.\,\ref{figiundgevecounts} whose axes were passing through the central detector ''carpet'' of particles detectors of the shower installation, with the underground detector placed beneath. More precisely, a formal rule was applied that the distance between the location of an EAS axis and the center of the ''carpet'' must be below 10\,m, which is comparable with the lateral size of the underground detector. Thus, like the case of the \HADRON experiment discussed in previous section the data gained presently relate to the central EAS region just around the shower axis.

The lower threshold of effective shower trigger elaboration in the considered experiment was set, as before, to the minimum EAS size of $N_e\simeq 10^5$.

By the middle of the year 2020 the whole duration of the period of simultaneous operation of the EAS particles detectors with the underground detector amounted to $\sim$$17000$~hours. During this time it was registered $\sim$$(0.9\cdot 10^6)$ showers which satisfied the above selection criterion. Among them of about $18000$~EAS events were accompanied by non-zero neutron response in the underground detector.
%


Relatively scarce statistics of the EAS with neutron events from muon interaction can be explained by the small geometrical factor of the underground detector which diminishes the probability for shower muons to hit its internals even at close axis location. Indeed, the total cross-section of this detector is of $\sim$$7.5$\,m$^2$ only, $20$ times less than the sensitive area of the muon hodoscope and XREC in the \HADRON installation. Moreover, predominant share of muons in EAS has the energy much below 1\,TeV, and correspondingly the average multiplicity of produced neutrons in the detector is essentially less then unit (see Fig.\,\ref{figiundgcalibri}). Nevertheless, since registration of EAS events was controlled exclusively by the trigger from the installation of particles detectors,
all showers with the size $N_e\geqslant 10^5$ were selected with equal probability, and the low detection efficiency of EAS muons by their neutron response did not influence the further statistical analysis of the data in considered experiment.

Splitting the whole statistics of neutron events in the underground detector into separate groups with different thresholds $M_T$ is equivalent to selection of the events which were induced by the muons with various minimum energy $E_\mu^{min}$. If to refer to the plot of Fig.\,\ref{figiundgcalibri}, it can be found that the sequence of the neutron multiplicity limits $M_T=(1, 2, 3, 5, 10)$ in Fig.\,\ref{figiundgevecounts} corresponds to a set of energy thresholds of muon detection $E_\mu^{min}\approx (25, 60, 100, 200, 450)$\,TeV, in supposition that the passage events of a  single muon through the underground detector must prevail at those high energies. Thus, in contrast to the hodoscope data of the \HADRON experiment, the underground detector potentially permits to study the muonic EAS content simultaneously in a number of energy ranges, and with much higher energy threshold of detected muons in general.

Alternatively, multiplicity of the detected neutron signals $M$ is evidently proportional to the intensity of the total flux of muons which traverse the underground detector in EAS events, and by simultaneous passage of a group of muons it is not necessarily required that all of them were of a TeV scale energy to produce noticeable neutron response. According to Fig.\,\ref{figiundgcalibri}, the expected average multiplicity of neutron signals for the muons with the energy of several GeV order is $(1-3)\cdot 10^{-3}$, so it is necessary to have $\sim$1000 muons at once hitting the neutron detector to achieve an assured registration of neutron event with $M\sim 1$. Taking into account the $7.5$\,m$^2$ area of the detector, the minimal flux of such muons must be of about $\Phi_\mu^{min}\simeq (100-150)$\,particles/m$^{2}$.

On the other hand, the value of particles flux $\Phi_\mu^{min}$ can be compared with the function $\varphi_{\mu}(r)$ which was mentioned in Section \ref{secmuohodo} and describes the lateral density distribution of the $\geqslant$5\,GeV EAS muons as measured by the old Tien Shan hodoscope. As it follows from Fig.~\ref{figixrec}, in the events with EAS sizes $N_e\simeq (10^5-10^7)$ the total amount of muons $N_\mu$ varies between the values of $(10^3-10^5)$, so in the distance range of $r\simeq (1-10)$\,m, \idest around the EAS center, the distribution function $\varphi_{\mu}(r)$ predicts the muon density of about $\varphi_{\mu}(r)\cdot N_\mu \simeq(1-500)$\,particles/m$^{2}$. This estimation agrees, in reasonable limits, with the minimum flux of EAS muons $\Phi_\mu^{min}$ which was deduced above from the neutron detector data.
In such interpretation the series of multiplicity thresholds $M_T$ in Fig.\,\ref{figiundgevecounts} corresponds to variation of the minimal density of low-energy muons in the central region of EAS which would be sufficient to generate an observed amount of neutron signals.

Since it was not possible to distinguish decidedly between the two alternatives of neutron signal origin at the present stage of the Tien Shan experiment, it would be more correct now to tell about the integral energy release of the penetrative EAS component  into evaporation neutrons when dealing with the data of the underground neutron detector. Satisfactory agreement between the numerical estimations of the muon flux $\Phi_\mu$ which result from the hodoscope and underground detector data means that at least a part of the neutron signals detected in the central EAS region must be due to a multitude of muons with relatively moderate energies of a few GeV, though admixture of small amount of particles with a TeV scale energy can not be excluded also. More on this subject follows below.

As it is seen in Fig.\,\ref{figiundgevecounts}, the dependencies of the relative share of showers accompanied by underground neutron events, ${N_{M\geqslant M_T}}\left/\right.{N_{EAS}}$, generally look similar to a power-like function of $N_e$ but experience sharp variation of the slope index around the point of $N_e\approx 10^6$, \idest near the position of the 3\,PeV knee in the primary cosmic ray spectrum. Since the original generation source of the detected neutron events is connected with EAS muons, the change revealed in behavior of the ${N_{M\geqslant M_T}}\left/\right.{N_{EAS}}$ relation means corresponding rise, at the same point of the spectrum knee, of the average multiplicity, or energy, or both of the muonic EAS component. To specify this change more quantitatively, the experimental distribution can be approximated by a piecewise power function of $N_e$ with combination of two different power indices:
\[
 F(N_e)=\left\{
\begin{array}{cl}
A\cdot N_e^{\alpha_1} & \mbox{if } N_e<N_e^\star, \\
A\cdot (N_e^\star)^{\alpha_1-\alpha_2}\cdot N_e^{\alpha_2} & \mbox{if otherwize.}
\end{array}
\right.
\]
Here, the parameter $N_e^\star\sim (0.3-1)\cdot 10^6$ is a border value of shower sizes where the distribution of the ${N_{M\geqslant M_T}}\left/\right.{N_{EAS}}$ relation experiences a kink, and $A$ is an arbitrary normalization coefficient. The four quantities $A$, $\alpha_1$, $\alpha_2$, and $N_e^{\star}$ can be considered as free parameters whose concrete values can be defined through fitting the experimental points in Fig.\,\ref{figiundgevecounts} by the standard $\chi^2$ minimization method.
Such approximating functions are plotted  by the thick broken lines in Fig.\,\ref{figiundgevecounts}, while the resulting best fit values for $N_e^{\star}$, $\alpha_1$, and $\alpha_2$, together with the variation $\Delta\alpha$ of power index at the kink point are listed in Table\,\ref{tabundgevecounts}. 

\begin{table*}
\begin{center}
\caption{Best fit parameters of the piecewise power functions $F(N_e)\sim N_e^{\alpha}$ which approximate the experimental distributions $N_{M\geqslant M_T}/N_{EAS}$ in Figure~\ref{figiundgevecounts}: position of the border point $N_e^\star$, and power indices both before ($\alpha_1$) and after ($\alpha_2$) the kink.}
\label{tabundgevecounts}
\begin{tabular*}{\textwidth}{@{\extracolsep{\fill}}c|ccccc|ccccc}
\hline
 $M_T$ & \multicolumn{5}{c|}{\textit{UPPER}} & \multicolumn{5}{c}{\textit{LOWER}} \\
&
{\small $N_e^\star/10^5$}&
{\small$\alpha_1$} &
{\small$\alpha_2$} &
{\small$\Delta\alpha$} &
{\small$\Delta\alpha/\alpha_1$} &
{\small $N_e^\star/10^5$} &
{\small$\alpha_1$} &
{\small$\alpha_2$} &
{\small$\Delta\alpha$} &
{\small $\Delta\alpha/\alpha_1$}  \\
\hline
1 &
4.7&
0.15 &
0.51 &
0.36 &
2.4&

5.9&
0.17 &
0.53 &
0.36 &
2.1
\\

2 &
3.4&
0.21 &
0.74 &
0.53 &
2.5&

4.3&
0.33 &
0.70 &
0.37 &
1.1
\\

3 &
3.4&
0.18 &
0.79 &
0.61 &
3.4 &

4.0&
0.37 &
0.78 &
0.41 &
1.1
\\

5 &
5.5&
0.24 &
0.91 &
0.67 &
2.8 &

7.0&
0.55 &
0.84 &
0.29 &
0.53
\\

10 &
5.6&
0.013 &
1.11 &
1.1&
84&

7.8&
0.37 &
0.77 &
0.40&
1.1
\\
\hline
\end{tabular*}
\end{center}
\end{table*}

According to the data of Table\,\ref{tabundgevecounts}, the relative change $\Delta\alpha/\alpha_1$ of power index in the upper detector unit generally varies between the limits of $2$ and $3$,
and experiences a sharp leap up to $\sim$$80$ in the last distribution with $M_T=10$.
In the lower unit this relation occurs to be essentially less, and remains of about $(0.5-1)$ for all distributions, with remarkable exception of the leading one, that with $M_T=1$, where the power indices $\alpha_1$ and $\alpha_2$ in the both units are nearly the same. The best fit values $N_e^{\star}$ of the kink position in all distributions vary between $(3.5-7)\cdot 10^5$, \idest all of them immediately precede the knee of the cosmic ray spectrum at $N_e\simeq 10^6$.

Systematic excess of the $\Delta\alpha/\alpha_1$ relation in the upper unit above the lower means that it is the EAS muon component with a moderate energy of several GeV order which is mostly responsible for the change of the ${N_{M\geqslant M_T}}\left/\right.{N_{EAS}}$ dependency slope in the vicinity of the $N_e\simeq 10^6$ point. Indeed, since the iron absorber separating the detector units is equivalent to addition of about $10$\,GeV to the muon registration threshold of the lower unit comparatively with the upper one, it can not cause any difference between the two units in regard to detection of the high-energy, TeV scale, muons. At the same time, this absorber essentially modifies the registration threshold of the lower unit relative to the muons in the GeV energy range. Consequently, the reduced $\Delta\alpha$ difference in the lower unit indicates that the low-energy part of the EAS muon component must play decisive role by excessive production of evaporation neutrons in the showers with $N_e\geqslant 10^6$.

Nevertheless, it is seen in Table\,\ref{tabundgevecounts} that both the values $\alpha_{1}$ and $\alpha_{2}$ themselves and the difference $\Delta\alpha$ between them are close in both units for the two distributions of the ${N_{M\geqslant 1}}\left/\right.{N_{EAS}}$ relation which were calculated with the minimum multiplicity threshold $M_T=1$.  It follows from this fact that particular amount of the high-energy muons with $E_\mu \gg 10$\,GeV must participate as well in origination of the discussed effect. Most probable energy estimate for that component would be the one which results from the plot of Fig.\,\ref{figiundgcalibri} for the neutron events with $M\simeq 1$, \idest $E_\mu\simeq (20-50)$\,TeV.

A sudden change revealed now in the dependency on $N_e$ of the average neutron deposit from the side of EAS muonic component in the central region of showers confirms qualitatively, but in a quite different manner, the former conclusion of the \HADRON experiment on systematic multiplicity growth of the flux of penetrative particles in the center of the showers which were selected by the presence of families in X-ray films. Another similarity between these heterogeneous peculiarities in behavior of EAS muons is that both of them start to be observable in the same range of primary EAS energies around the 3\,PeV knee in the energy spectrum of cosmic rays.

\section{Conclusion}

By experimental investigation of the muon component of EAS in the range of primary cosmic ray energies $(1-100)$\,PeV at the Tien Shan mountain station the results were acquired which can be summarized as the following:
\begin{itemize}
\item[-]
Actually, the behavior of the low-energy, $E_\mu\geqslant 5$\,GeV, muon component of EAS is quite opposite to what should be intuitively expected starting from the naive supposition on  the production mechanism of gamma-hadron families in X-ray films. Namely, if to select the EAS events by presence of a family in the Tien Shan X-ray emulsion chamber, the multiplicity of muon component in the selected showers occurs larger in average than the mean muon number calculated over the whole EAS set. At the same time, from general models of hadronic interaction it follows that families must be preferably born in the showers originating from the light nuclei, so the muonic content in such selection should be essentially below the average.

\item[-]
The effect of superfluous, in comparison with the average, abundance of the muon component in the events with families starts to be observable in the showers with the primary energy of about $3$\,PeV, \idest at the knee of the cosmic ray spectrum, and spatially this effect is bound to the central region of EAS at a $\lesssim $$1$\,m distance from shower axis.

\item[-]
Acceleration of the relative transmission of EAS energy into the muon component in shower cores was traced as well near the same border point of $(1-3)$\,PeV in the data on multiplicity of evaporation neutrons produced by muons in absorber of the Tien Shan underground detector. The effect of enhanced neutron production results mostly from interaction of the EAS muons with moderate energy $E_\mu$ of about $(5-30)$\,GeV, though a noticeable share in its origin comes also on the part of the high-energy muonic component with $E_\mu\simeq (20-50)$\,TeV.

\end{itemize}

Qualitative correspondence between the analogous conclusions of two independent experiments made at Tien Shan on enhanced rise of muon flux in the EAS around the knee point of cosmic ray spectrum is an evidence of sufficient reliability of both detected effects. As well, the results considered in present publication correspond to the lately found ''muon puzzle'' tendency of peculiarly high muon production by cosmic rays around and above the $\sim$$10$\,PeV region of their energy spectrum. As known, this phenomenon, actively discussed in current literature, contradicts both to contemporary models of hadron interaction, and to modern views on the atomic composition of primary cosmic ray nuclei
\cite{muons2,bacsan1997,cern_muons_aleph_2003,cern_muons_delphi_2007,petrukhin_puzzle_2014,kascade_muons_2018,auger_hadron_probing_2020}.

\section*{Acknowledgements}

This work was partially supported by the grant \#AP09258896 of the Ministry of Education and Science of Republic of Kazakhstan.

%
%
%
\newcommand {\etal}{\textit{et al.}}

\end{document}